\documentclass[twocolumn,twoside,showpacs,preprintnumbers,amsmath,amssymb,prd,superscriptaddress,floatfix]{revtex4}

\usepackage{graphicx}
\usepackage{dcolumn}
\usepackage{array}
\usepackage{color}
\usepackage{longtable}
\usepackage{multirow}
\usepackage{bm}

\begin{document}

\preprint{APS/123-QED}

\newcommand*{\INFNRoma}{Istituto Nazionale di Fisica Nucleare, Sezione di Roma1, P.le A.Moro 2, I-00185, Roma, Italy}
\affiliation{\INFNRoma}
\newcommand*{\Frascati}{Laboratori Nazionali di Frascati, Istituto Nazionale di Fisica Nucleare, Via E.Fermi 40, I-00044, Frascati, Italy}
\affiliation{\Frascati}
\newcommand*{\Trento}{Dipartimento di Fisica, Universit\`a di Trento, I-38050 Povo, Trento, Italy}
\affiliation{\Trento}
\newcommand*{\TrentoINFN}{Istituto Nazionale di Fisica Nucleare, Gruppo Collegato di Trento, Sezione di Padova, I-38050 Povo, Trento, Italy}
\affiliation{\TrentoINFN}
\newcommand*{\TVergata}{Dipartimento di Fisica, Universit\`a di Roma ``Tor Vergata'', Via Ricerca Scientifica 1,  I-00133 Roma, Italy}
\affiliation{\TVergata}
\newcommand*{\RomaDue}{Istituto Nazionale di Fisica Nucleare, Sezione di RomaDue, Via Ricerca Scientifica 1,  I-00133 Roma,Italy}
\affiliation{\RomaDue}
\newcommand*{\UWA}{Department of Physics, University of Western Australia, Nedlands, WA 6907 Australia}
\affiliation{\UWA}
\newcommand*{\CeFSA}{Centro di Fisica degli Stati Aggregati, IFN-CNR, Trento, I-38050 Povo, Trento, Italy}
\affiliation{\CeFSA}
\newcommand*{\CNR}{Istituto Fisica Spazio Interplanetario, CNR, Via Fosso del Cavaliere, I-00133 Roma, Italy}
\affiliation{\CNR}
\newcommand*{\LSU}{Department of Physics and Astronomy, Louisiana State University, Baton Rouge, Louisiana 70803}
\affiliation{\LSU}
\newcommand*{\lAquila}{Dipartimento di Fisica, Universit\`a de L'Aquila, and INFN, L'Aquila,Italy}
\affiliation{\lAquila}
\newcommand*{\Padova}{Dipartimento di Fisica, Universit\`a di Padova, Via Marzolo 8, 35131 Padova, Italy}
\affiliation{\Padova}
\newcommand*{\PadovaINFN}{Istituto Nazionale di Fisica Nucleare, Sezione di Padova, Via Marzolo 8, 35131 Padova, Italy}
\affiliation{\PadovaINFN}
\newcommand*{\LaSapienza}{Dipartimento di Fisica, Universit\`a di Roma ``La Sapienza'', P.le A.Moro 2, I-00185, Roma, Italy}
\affiliation{\LaSapienza}
\newcommand*{\Ferrara}{Dipartimento di Fisica, Universit\`a di Ferrara, I-44100 Ferrara, Italy}
\affiliation{\Ferrara}
\newcommand*{\FerraraINFN}{Istituto Nazionale di Fisica Nucleare, Sezione di Ferrara I-44100 Ferrara, Italy}
\affiliation{\FerraraINFN}
\newcommand*{\MIT}{Whitehead Institute, MIT, Cambridge, Massachusetts, USA}
\affiliation{\MIT}
\newcommand*{\Loyola}{Department of Physics, Loyola University, New Orleans, Louisiana, USA}
\affiliation{\Loyola}
\newcommand*{\LNL}{Laboratori Nazionali di Legnaro, Istituto Nazionale di Fisica Nucleare, 35020 Legnaro, Padova, Italy}
\affiliation{\LNL}
\newcommand*{\IESS}{Istituto Elettronica Stato Solido, IFN-CNR, Via Cineto Romano 42, Roma, Italy}
\affiliation{\IESS}

\title{Methods and results of the IGEC search for burst gravitational waves\\ in the years 1997--2000}
\author{P.~Astone}
\affiliation{\INFNRoma}
\author{D.~Babusci}
\affiliation{\Frascati}
\author{L.~Baggio}
\affiliation{\Trento}
\affiliation{\TrentoINFN}
\author{M.~Bassan}
\affiliation{\TVergata}
\affiliation{\RomaDue}
\author{D.G.~Blair}
\affiliation{\UWA}
\author{M.~Bonaldi}
\affiliation{\CeFSA}
\affiliation{\TrentoINFN}
\author{P.~Bonifazi}
\affiliation{\CNR}
\affiliation{\INFNRoma}
\author{D.~Busby}
\affiliation{\LSU}
\author{P.~Carelli}
\affiliation{\lAquila}
\author{M.~Cerdonio}
\affiliation{\Padova}
\affiliation{\PadovaINFN}
\author{E.~Coccia}
\affiliation{\TVergata}
\affiliation{\RomaDue}
\author{L.~Conti}
\affiliation{\Padova}
\affiliation{\PadovaINFN}
\author{C.~Cosmelli}
\affiliation{\LaSapienza}
\affiliation{\INFNRoma}
\author{S.~D'Antonio}
\affiliation{\Frascati}
\author{V.~Fafone}
\affiliation{\Frascati}
\author{P.~Falferi}
\affiliation{\CeFSA}
\affiliation{\TrentoINFN}
\author{P.~Fortini}
\affiliation{\Ferrara}
\affiliation{\FerraraINFN}
\author{S.~Frasca}
\affiliation{\LaSapienza}
\affiliation{\INFNRoma}
\author{G.~Giordano}
\affiliation{\Frascati}
\author{W.O.~Hamilton}
\affiliation{\LSU}
\author{I.S.~Heng}
\altaffiliation[presently at ]{Max-Planck-Inst. f\"ur Gravitationsphysik,
Albert-Einstein-Inst. Hannover,
Callinstr. 38,
30167 Hannover,
Germany
}
\affiliation{\LSU}
\author{E.N.~Ivanov}
\affiliation{\UWA}
\author{W.W.~Johnson}
\affiliation{\LSU}
\author{A.~Marini}
\affiliation{\Frascati}
\author{E.~Mauceli}
\affiliation{\MIT}
\author{M.P.~McHugh}
\affiliation{\Loyola}
\author{R.~Mezzena}
\affiliation{\Trento}
\affiliation{\TrentoINFN}
\author{Y.~Minenkov}
\affiliation{\RomaDue}
\author{I.~Modena}
\affiliation{\TVergata}
\affiliation{\RomaDue}
\author{G.~Modestino}
\affiliation{\Frascati}
\author{A.~Moleti}
\affiliation{\TVergata}
\affiliation{\RomaDue}
\author{A.~Ortolan}
\affiliation{\LNL}
\author{G.V.~Pallottino}
\affiliation{\LaSapienza}
\affiliation{\INFNRoma}
\author{G.~Pizzella}
\affiliation{\TVergata}
\affiliation{\Frascati}
\author{G.A.~Prodi}
\email[Corresponding author ]{prodi@science.unitn.it}
\affiliation{\Trento}
\affiliation{\TrentoINFN}
\author{L.~Quintieri}
\affiliation{\Frascati}
\author{A.~Rocchi}
\affiliation{\TVergata}
\author{E.~Rocco}
\affiliation{\Trento}
\author{F.~Ronga}
\affiliation{\Frascati}
\author{F.~Salemi}
\affiliation{\Ferrara}
\author{G.~Santostasi}
\affiliation{\LSU}
\author{L.~Taffarello}
\affiliation{\PadovaINFN}
\author{R.~Terenzi}
\affiliation{\CNR}
\author{M.E.~Tobar}
\affiliation{\UWA}
\author{G.~Torrioli}
\affiliation{\IESS}
\author{G.~Vedovato}
\affiliation{\LNL}
\author{A.~Vinante}
\affiliation{\Trento}
\affiliation{\TrentoINFN}
\author{M.~Visco}
\affiliation{\CNR}
\affiliation{\RomaDue}
\author{S.~Vitale}
\affiliation{\Trento}
\author{J.P.~Zendri}
\affiliation{\PadovaINFN}

\collaboration{International Gravitational Event Collaboration}
\homepage{http://igec.lnl.infn.it/}
\noaffiliation

\date{\today}

\begin{abstract}
This paper presents the results of the observations of the detectors participating in the International Gravitational Event Collaboration (IGEC) from 1997 to 2000 and reviews the data analysis methods.  The analysis is designed to search for coincident excitations in multiple detectors.  The data set analysed in this article covers a longer period and is more complete than that given in previous reports.  The current analysis is more accurate for determining the false dismissal probability for a time coincidence search and it optimizes the search with respect to a target amplitude and direction of the signal. The statistics of the accidental coincidences agrees with the model used for drawing the results. The observations of this  IGEC search are consistent with no detection of gravitational wave burst events.  A new conservative upper limit has been set on the rate of
gravitational wave bursts with Fourier component $H > 2\cdot 10^{-21}Hz^{-1}$, both for searches with and without a filter for the Galactic Center direction.  This study confirms that
the false alarm rate of the observation can be negligible when at least
three detectors are operating simultaneously.

\end{abstract}

\pacs{0480Nn, 9585Sz}
\maketitle

\section{\label{sec:level1}Introduction}

This paper presents the results of the observations of the International Gravitational Event Collaboration (IGEC) from 1997 to 2000.  We have made an extensive search for burst-type gravitational waves with the largest network of detectors ever assembled, and report here the details of the search and the new upper limits achieved for the rate of gravitational wave burst events.

The search for gravitational waves (gw) involves detecting the presence of a signal in the noise of the detector array.  A signal must compete with the intrinsic noise of the detectors, and also with transient excitations (of mechanical or electromagnetic origin for example) which usually can not be discriminated from the actual gw signal. Therefore, it is not viable to perform burst gw searches with a single detector. With two or more detectors in simultaneous observation, the impact of local transient excitations on burst gw searches is significantly reduced. Moreover, the false alarm rate can be reliably estimated. Hence, to facilitate multi-detector searches for burst gw, the International Gravitational Event Collaboration (IGEC) was formed in 1997 \footnote{See http://igec.lnl.infn.it}\newcounter{link}\setcounter{link}{\value{footnote}}. This collaboration currently consists of five cryogenic resonant-bar gravitational wave detectors, ALLEGRO~\cite{ALLEGRO}, AURIGA~\cite{AMA2GAP}, EXPLORER~\cite{EXPLORER}, NAUTILUS~\cite{NAUTILUS} and NIOBE~\cite{NIOBE}, operating as a worldwide network. The members of this network exchange lists of candidate gw events and related information under an agreed data exchange protocol. 

The target signals are transients without structure
in the frequency range investigated. 
Examples of such signals are short pulses of $ \sim 1\ ms$ duration,
signals showing a few cycles of $\sim 1\ ms$ period and signals sweeping in
frequency across $\sim 1\ kHz$.  Possible sources are therefore related
to compact astrophysical objects, like the coalescence of neutron star
and black hole binaries \cite{sources1,sources2}.

The main method used to search for burst gw has been to search for an
excess of coincident excitations in two detectors \cite{Amaldi89,ALEX99,Coinc1,Coinc2}. The
IGEC performed the first thorough search on more than two detectors on
data acquired in 1997 and 1998  \cite{IGECPRL}. No claims of detection
were made, but an improved upper limit on burst gw was set.  In 2001 all
the data acquired by IGEC members between 1997 and 2000 were
exchanged.  A preliminary search was performed on these data \cite{IGECCQG2002}.

In this article, we present the results of a comprehensive search by the IGEC for burst gravitational waves on the full data set.  In addition to the extended observation time, this analysis makes significant progress over the previous searches in the following respects: (i) different search thresholds are systematically tried; (ii) the time coincidence window is determined by the desired confidence level; (iii) a directional search strategy is implemented; (iv) the statistics of the estimated false alarms is thoroughly investigated and (v) the statistical methods chosen to set the confidence intervals ensure a given \textit{coverage}~\footnote{In this analysis, we refer to the conventional frequentist \textit{coverage}, i.e. the probability that the confidence interval contains the true gw value. In other words, the coverage is the probability as could be measured in principle by repeating the same observations with the same gw source, where ``same" is meant in the stochastic sense. In this analysis, we made conservative estimates of the coverage.}, i.e. the probability that the confidence interval contains the true value.  In the following Section, we review the IGEC operation during 1997-2000. 
The methods of the multi detector analysis are described in Section~\ref{sec:multidet}.
Finally, in Section~\ref{sec:results}, we discuss the results. In this section we pay particular attention to the new upper limit on the rate of detected burst gw~\footnote{The authors P.Astone, G.V. Pallottino and G. Pizzella point out that the analysis presented in this paper is aimed at systematic search for coincident excitations in multiple detectors, however they are convinced that a probabilistic estimation of the flux of g.w. on Earth as well as of upper limits requires a Bayesian approach.} and on the low level of
false alarms achieved by this observatory.

\begin{table*}[!ht]
\caption{\label{tab:dets}Summary of detector characteristics. The reported misalignment is the angle between the bar axis and a common direction. The observation time refers to the data exchanged for this 1997-2000 IGEC analysis.}
\begin{ruledtabular}
\begin{tabular}{cccccc}
Detector&	ALLEGRO&	AURIGA&	EXPLORER&	NAUTILUS&	NIOBE\\ \hline
Material&	Al5056&	Al5056&	Al5056&		Al5056&	Nb\\ \hline
Mass [kg] &	2296&	2230&	2270&	2260&	1500\\ \hline
Length [m]&	3.0&	2.9&	3.0&	3.0&	2.8\\ \hline
\multirow{2}{*}{Resonant frequencies [Hz]}&	920&	930&	921&	924&	713\\ \cline{2-6}
&	895&	912&	905&	908&	694\\ \hline
Temperature [K]&	4.2&	0.2&	2.6&	0.1&	5.0\\ \hline
Longitude&	$268^\circ$50'E&	$11^\circ$56'54''E&	$6^\circ$12'E&	$12^\circ$40'21''E&	$115^\circ$49'E\\ \hline
Latitude&	$30^\circ$27'N&	$45^\circ$21'12''N&	$46^\circ$27'N&	$41^\circ$49'26''N&	$-31^\circ$56'N\\ \hline
Azimuth&	$-40^\circ$E&	$44^\circ$E&	$39^\circ$E&	$44^\circ$E&	$0^\circ$\\ \hline
Misalignment [deg]&	9&	4&	2&	3&	29\\ \hline
Observation time [d]&	852.5&	216.5&	551.0&	414.8&	192.6
\end{tabular}
\end{ruledtabular}
\end{table*}

\section{The exchanged data}

In this Section we review the 1997-2000 operation of the IGEC. We recall the sensitivity to burst gws of the participating bar detectors (see~\ref{sec:detectors}). Each detector group searches its data independently for \emph{gravitational wave candidates}, or \emph{events}. Then, the information exchanged under IGEC is described, with particular attention to the data validation requirements (see~\ref{sec:protocol}). The quality of the contribution of each detector to the IGEC observatory is discussed in terms of observation time and false alarms as a function of a threshold on the amplitude of target gw signals (see~\ref{sec:data}). 
The statistics of the time series of the exchanged events show autocorrelation at short timescales. This clustering however disappears when cross-correlating different detectors (see~\ref{sec:times}).

\subsection{The detectors}
\label{sec:detectors}

All of the currently operating resonant detectors measure the tidal strain of a
mechanically isolated cylindrical bar caused by impinging gws.  A lighter mechanical resonator
is strongly coupled and tuned to the fundamental longitudinal mode of the bar, resulting in a
system with two normal modes of vibration. 

A list of the main characteristics of the detectors is shown in Table~\ref{tab:dets}. All the bars are cooled to cryogenic temperatures. AURIGA and NAUTILUS operate at a few hundred $mK$ to further reduce contributions from the thermal noise.  For the same reason, Niobium was chosen as the material for NIOBE because it has a higher mechanical quality factor at 4K.  

The IGEC search is focused on burst gws, which can be modeled as a pure Dirac $\delta$-function excitation. The strength of a burst can be quantified by its Fourier amplitude $H_o$, or \emph{amplitude}, which is related to the energy $E_s$ deposited in the bar by 
\begin{equation}
H_0  = \frac{1}{{4L\nu _o^2 }}\sqrt {\frac{{E_s }}{M}}
\label{eq:H0}
\end{equation}
where $L$ is the bar length, $M$ its mass, $\nu_0$ the mean resonant frequency of the detector.

Actually, the class of detectable signals is much wider than $\delta$-functions. This search is also effective for all short duration signals which have an almost constant value for their Fourier amplitude at the detector frequencies, i.e. $\sim 700~Hz$ for NIOBE and $\sim 900~Hz$ for the other detectors (see Table~\ref{tab:dets}). In these cases, the gw amplitude $H_o$ is estimated without bias. 

The detectors were oriented to be nearly parallel to each other.  This was done by orienting them to be perpendicular to a common great circle that passes through or near the sites.  This makes their antenna patterns coherent and maximizes the probability of coincident signal detection between multiple detectors.

\begin{figure}
\centering
\includegraphics[width=85mm]{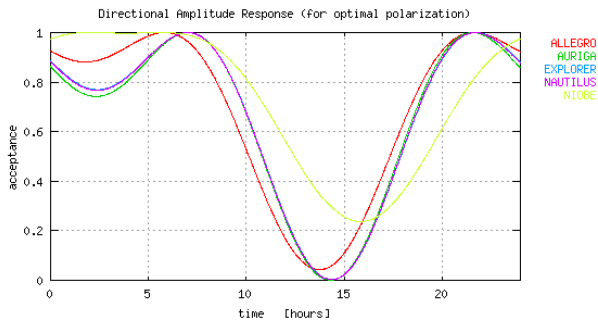}
\caption{\label{fig:pattern}Amplitude directional sensitivity of the detectors versus UTC with respect to the Galactic Center for DAY 25 Dec 2000.}
\end{figure}

For cylindrical bar detectors, the amplitude observed for a gw signal
from a particular source in the sky follows a $\sin^2 \theta$ function,
where $\theta$ is the angle between the long axis of the bar and the
direction of the source. As a demonstration of the effectiveness of
the chosen common orientation, we plot in Figure~\ref{fig:pattern} the directional sensitivity with
respect to the Galactic Center for the five detectors of the IGEC.  The
sensitivity to wave polarization is then $\cos 2 \psi$, where $\psi$
is the polarization angle in the wavefront plane with respect to the
projection of the bar axis.

\begin{figure*}
\centering
\includegraphics[width=175mm]{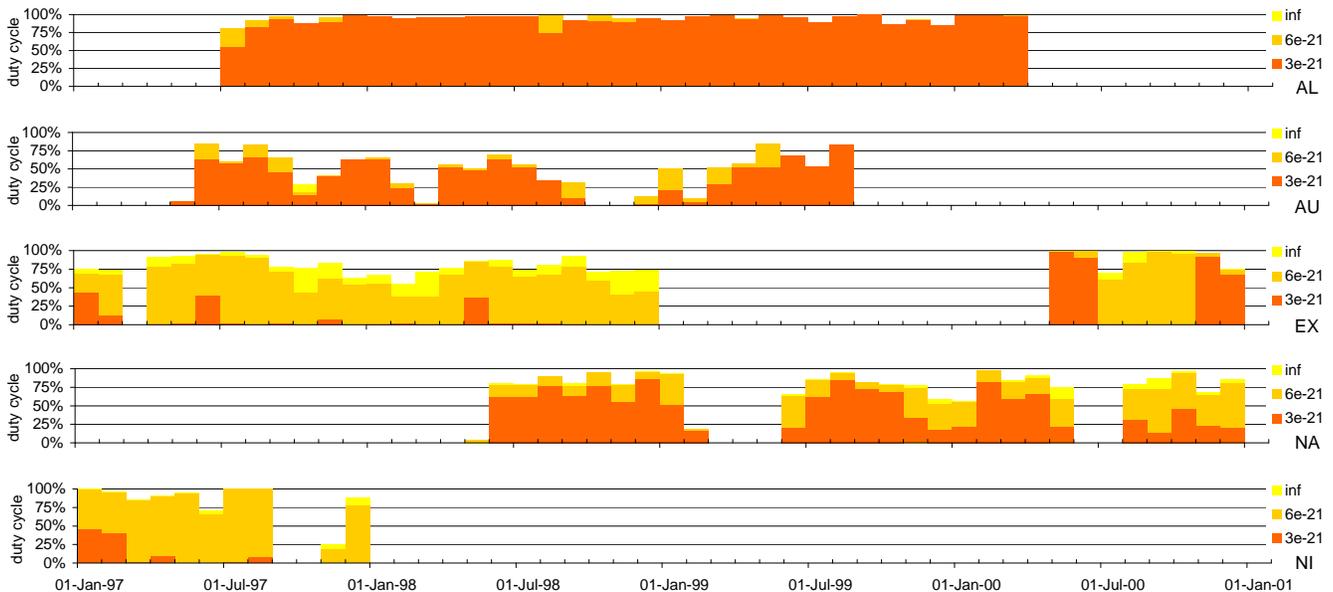}
\caption{\label{fig:onoff}Overview of the observation periods in 1997-2000 for each detector, as fractions of time in monthly bins. Three different ranges of sensitivities are considered: exchange threshold lower than $3\cdot10^{-21}~Hz^{-1}$ (\textit{darker shade}), included in $3 - 6\cdot10^{-21}~Hz^{-1}$ (\textit{middle shade}) and above $6\cdot10^{-21}~Hz^{-1}$ (\textit{light shade}).}
\end{figure*}

\begin{figure}[t]
\centering
\includegraphics[width=85mm]{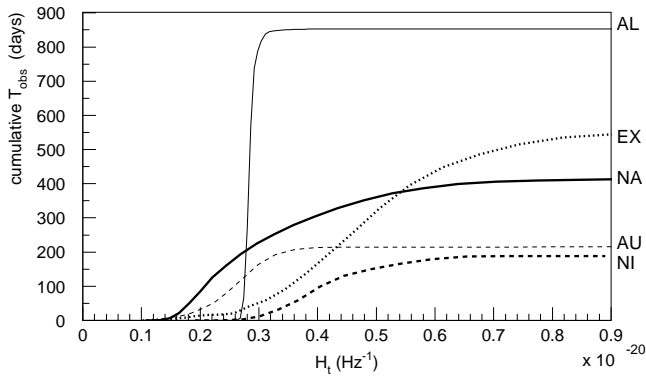}
\caption{\label{fig:obstime}Observation time of the IGEC detectors as a function of a threshold $H_t$ on gw amplitude in 1997-2000. The ordinate is the integrated time during which the detector exchange threshold has been lower than $H_t$.}
\end{figure}

\subsection{The IGEC data exchange protocol}
\label{sec:protocol}
Each group implements independently a burst gw search by using an optimal filter 
\cite{McDonough}, which takes into account the slow variations of the noise characteristics of the detector~\cite{ALLEGRO, ROGAnalisys2, CHI2}. The filter estimates the Fourier amplitude of the burst gw. An adaptive threshold, \emph{exchange threshold}, is then applied to the filtered data and a list of the \emph{candidate events} above threshold is compiled \footnote{The time of arrival of the impulsive excitations is not known, therefore the optimal filter is applied continuously. The local maxima in the absolute value of the filtered output are compared to the threshold, and any maxima above are defined as candidate events. This modification of the Wiener theory introduces some non-linearity of the estimates. The resulting bias on the amplitude and time of arrival the events has to be taken into account. In the limit of high SNR events, linearity is recovered. See for instance Ref.~\onlinecite{AMA4Ortolan} for a discussion on the biases introduced at low SNR.}\newcounter{bias}\setcounter{bias}{\value{footnote}}. The exchange threshold is typically set to amplitude signal-to-noise ratio (SNR) between 3 and 5. The candidate events are described by the peak amplitude of the optimal filter output, the time of its occurrence and the uncertainties on amplitude and time.

The IGEC protocal requires each detector to report all time intervals of satisfactory operation. This is accomplished by vetoing periods corresponding to times of laboratory activity which are known to affect the sensitivity of the detector, such as periods of cryogenic maintenance. In the case of EXPLORER and NAUTILUS, times were also vetoed when the noise variance exceeded a certain value.  For AURIGA, times were vetoed when either the statistics of the noise was not Gaussian, or the Wiener filter was not properly matched to the noise\cite{AMA3GAP,AMA4Ortolan}.

All this information is exchanged within IGEC under a common data format. This protocol was last
updated in 2000\footnotemark[\value{link}]. 
The most relevant additions introduced by that update were: i) absence of any biases in the estimates of the time of arrival (\emph{ETA}) and \emph{ amplitude} \footnotemark[\value{bias}], ii) an estimate of the errors in ETA, iii) an upper bound of the systematic errors in amplitude, and iv) a continous measurement of the noise level and a continous record of the chosen exchange threshold (i.e. the threshold used to compile the event list). The noise level is described by the standard deviation and by the 3rd and 4th order moments of the noise distribution. 

The choice of the most suitable exchange threshold is left to each group, provided that two constraints are met.  One lower bound on the exchange threshold comes from the requirement of having unbiased estimates of the amplitude and ETA of the exchanged events. The other constraint is a specified criterion that limits the rate of exchanged events, and thus the rate of false alarms of the observatory.

The search for time coincidences among events of different detectors is performed by setting variable time windows computed from the ETA uncertainties to ensure a certain probability of false dismissal (see Section \ref{coincidences}). A rough estimate of the contribution of each individual detector to the final rate of accidental coincidences of the observatory can be found by considering the product of its average ETA standard deviation $\overline{\sigma_t}$ and its average event rate $\overline{\lambda}$ \footnote{See for instance Eq.1 of Ref.~\onlinecite{IGECPRL}, valid in the simple case of equal time uncertainties and uncorrelated noise performances. A more general discussion can be found in Ref.~\onlinecite{IGECtoolbox2001}}. The IGEC recommendation is that the threshold be kept high enough so that 
\begin{equation}
\label{eq:dtdl}
\overline {\sigma_t}  \times \overline {\lambda} = \frac{{\sum\nolimits_{i = 1}^n {\sigma_{t_i} } }}{{T_{obs} }} < 0.1\%
\end{equation}
where ${\sigma _t}_i $ is the standard deviation of the arrival time
for the $i$th event, $n$ is the {\it total} number of events and
$T_{obs}$ is the {\it total} observation time. This recommendation is an improvement of the one followed in the previous data exchange, i.e. to limit the rate to one hundred events per day~\cite{IGECGWDAW2000}, in order to cope with the  widening of the effective bandwidth of the detectors.

The ETA standard deviation has been estimated by means of a Monte Carlo simulation for the AURIGA, EXPLORER and NAUTILUS detectors \cite{AMA4Ortolan, ROGtimingPRD} or by measuring the response of the detector to repeated impulsive excitations for the ALLEGRO detector \cite{ALLEGRO}. The uncertainty on the ETA depends both on the noise level and on the timing accuracy of the filtered data. Therefore, the behavior of ${\sigma _t}$ is significantly different for the different detectors, though it decreases in all of them as SNR increases. It was also found to vary with time in AURIGA, EXPLORER and NAUTILUS, following the variations of the effective bandwidth of the detector \cite{IGECtoolbox2001, ROGtimingPRD}. Typical values of $\overline{\sigma_t}$ has been fractions of a second.

Whenever environmental monitors have been operating, the events have been checked against periods of ambient disturbances prior to their exchange. If an event from the filtered data occurs in coincidence with an excitation observed by these monitors, it is vetoed and not considered a candidate gw event. For AURIGA, no veto based on environmental monitor has been implemented, yet the event has to pass a $\chi^2$ test to check its consistency with an impulsive mechanical excitation of the bar\cite{CHI2}. This test has been found to provide a good reduction of false alarms at high SNR. Despite all such efforts to remove local excitations in the detectors, most of the events above threshold cannot be ruled out as candidate gw events \cite{Heng96}. Partial information about vetoed events are tracked in the exchanged event lists for diagnostic purposes.\footnote{For instance, the vetoed events have an associated dead time which have to be taken into account when computing the false dismissal of the observatory. In practice this amounts to removing a negligible fraction of the observation time.}

\begin{figure*}[!htb]
\includegraphics[width=175mm]{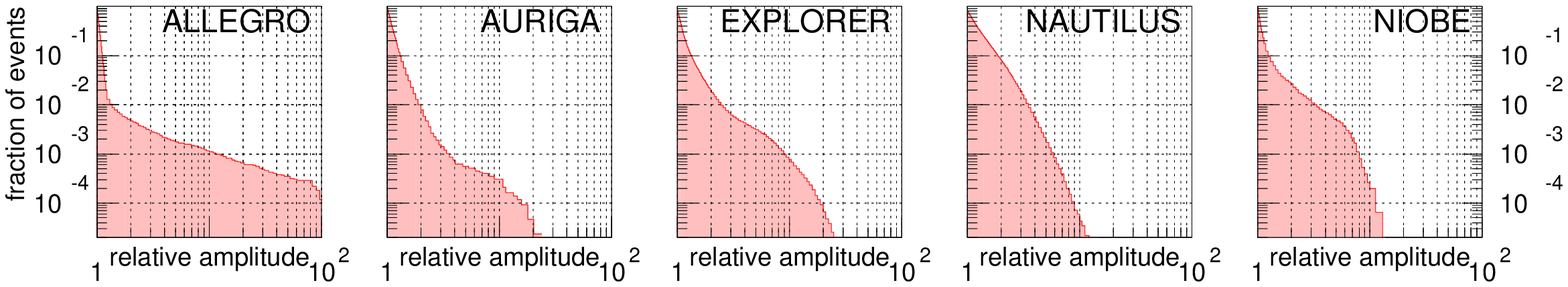}
\caption{\label{fig:amplitude}Cumulative distribution function of the amplitude of the exchanged events: the ordinate is the fraction of events whose amplitude is greater than the value in abscissa. The amplitude is normalized to the corresponding exchange threshold of the detector.
}
\end{figure*}

\begin{figure}[t]
\centering
\includegraphics[angle=270,width=85mm]{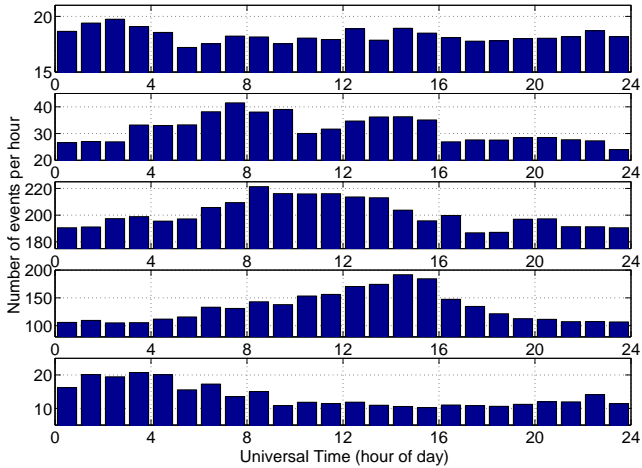}
\caption{\label{fig:hourlyrate}Average rate of exchanged events per each UTC hour
of the day. From top to bottom: ALLEGRO, AURIGA, EXPLORER, NAUTILUS and NIOBE.}
\end{figure}

\subsection{Data overview}
\label{sec:data}

The duty cycle of the network has been previously reported in detail \cite{IGECGWDAW2000,IGECCQG2002}. A graphical representation of the on-off times of the individual detectors between 1997 and 2000 is shown in FIG.~\ref{fig:onoff}. During this 4 year period, there were 1319 days when at least 1 detector was operating, 707 days with at least 2 detectors in simultaneous operation, 173 days with at least 3 detectors and 26 days with at least 4 detectors. This time coverage is a consequence of the non optimal overlap among the operating times of the single detectors and of their duty cycles, which were included within the 0.9 achieved by ALLEGRO and the 0.3 of AURIGA.

The cumulative observation time of each detector versus gw amplitude is shown in FIG.~\ref{fig:obstime}. 
The exchange thresholds of the detectors fluctuated significantly in time for all detectors but ALLEGRO, following the non stationary behavior of the noise. 

The typical range of the exchanged thresholds was  $2\div 6\cdot10^{-21}~Hz^{-1}$. Neglecting the directional sensitivity of the detector, this range corresponds to a $1ms$  burst generated by $0.02\div 0.2 M_\odot$ solar masses converted in gw with isotropic emission at a distance of $10kpc$. With respect to the first IGEC data exchange \cite{IGECGWDAW2000}, EXPLORER and NAUTILUS take the opportunity of the relaxed recommendation on the average rate of exchanged events (Eq.~\ref{eq:dtdl}), and lower the exchange threshold (from $SNR\approx 5$ to about 4.5 ). The data of the other detectors keep the previous exchange thresholds ($SNR\approx 3$ for ALLEGRO and NIOBE, $SNR\approx 5$ for AURIGA).
The mean rate of events exchanged by EXPLORER and NAUTILUS is about 5 times greater than in the previous exchange, while for the other detectors it remains at the same level.

The amplitude distributions of the exchanged events are shown in FIG.~\ref{fig:amplitude}. At least two distinct regimes are observed in all detectors, though showing a large variability among them: a steep roll-off close to the threshold, and an additional tail dominating at SNR greater than $\sim 10$. The exchanged events are mostly generated by non-Gaussian noise sources, which are not currently modeled.

The average number of events observed in each hour (Universal Time) is shown for the five detectors in FIG.~\ref{fig:hourlyrate}. All detectors show an increase in the number of events observed at certain hours of the day. For AURIGA, EXPLORER and NAUTILUS, higher event rates are observed between 5 and 18 hours local time, while for NIOBE between 7 and 16 hours local time. Such behavior is consistent with a correlation with human activity. The event rate for ALLEGRO is almost constant and shows a small increase between 18 and 23 hours local time. Note that since AURIGA, EXPLORER and NAUTILUS are in the same time zone, the rise and fall of event rates is almost synchronized.

\begin{figure}[!t]
\centering
\includegraphics[width=85mm]{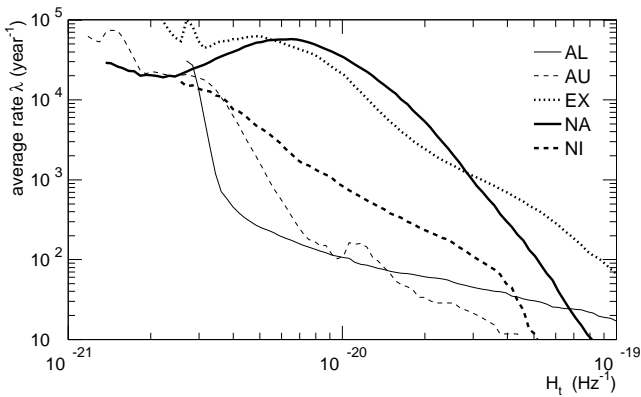}
\caption{\label{fig:meanrate}Average rate of the events exchanged by the detectors as a function of a threshold $H_t$ on gw amplitude, as described in the text. The minimum threshold plotted for each detector corresponds to 10 days of observation time.}
\end{figure}

The average rates of the events exchanged by each detector are shown in FIG.~\ref{fig:meanrate} as a function of an absolute \textit{search threshold} on the gw amplitude. The search threshold is applied to each event list before performing the coincidence search described in Section~\ref{coincidences}. For a given search threshold value $H_t$, the rate is computed dividing the number of events exceeding $H_t$  by the observation time during which the exchange threshold of the detector has been lower than $H_t$. This procedure is consistent with the data selection we applied in the multi detector analysis, as described in Section~\ref{sec:preprocessing}. In general, the event rates decrease as the search threshold increases because the number of selected events decreases and the selected observation time increases. However, this can happen in a non monotonic way due to the non-stationary behavior of the noise performances of the detectors. In fact, as the search threshold increases, the selected observation time can extend to additional periods of operation characterized by worse sensitivity. These periods contribute with a much higher instantaneous rate of events, given that most events appear very close to the exchange threshold (see FIG.~\ref{fig:amplitude}). Therefore, the mean event rate may increase at some higher search threshold.

\begin{figure}[!t]
\centering
\includegraphics[width=85mm]{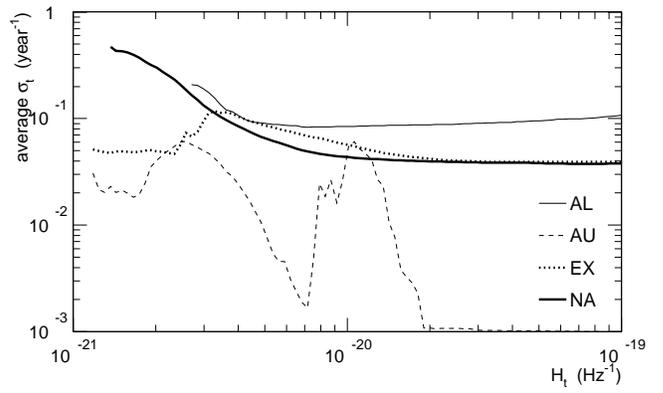}
\caption{\label{fig:meanMT2}Average standard deviation of the ETA of exchanged events, $\overline {\sigma_t}$, as a function of a threshold $H_t$ on gw amplitude, as described in the text. The minimum threshold plotted for each detector corresponds to 10 days of observation time. For NIOBE only a $1s$ upper limit for the time standard deviation is available.}
\end{figure}

The mean timing uncertainty, $\overline {\sigma_t}$, of the exchanged events is shown in FIG.~\ref{fig:meanMT2} for each detector as a function of the search threshold. The value of $\overline {\sigma_t}$ is dominated by the selected events which are closest to the exchange threshold of the detector. As the threshold increases, the mean timing uncertainty decreases down to the limit given by the timing calibration or resolution of the detectors, which is in the range $1\div 80\ ms$. Similarly to FIG.~\ref{fig:meanrate}, $\overline {\sigma_t}$ is a non-monotonic function of the search threshold due to noise being non-stationary. For instance, the peak shown by AURIGA data around $10^{-20}Hz^{-1}$ is due to a one week period of operation of the detector with reduced sensitivity and reduced effective bandwidth. For the NIOBE data a conservative estimate of $\overline {\sigma_t}$ was provided, independent from the event amplitude.

FIG.~\ref{fig:meanrate} and FIG.~\ref{fig:meanMT2} allow the reader to compare the contributions of each detector to the false alarm rate of the observatory as a function of the selected threshold $H_t$ on gw amplitude. In this respect, the quality of the detector is given by the product of the mean event rate and the mean uncertainty of the ETA, as already discussed in connection with eq.~\ref{eq:dtdl}. The cleanest detectors have been ALLEGRO and AURIGA in the investigated range of gw amplitudes. To be precise, the actual false alarm rate of the observatory is not directly related to the mean event rates. Rather it is time by time proportional to the product of the instantaneous event rates of the participating detectors \cite{IGECtoolbox2001}. Moreover, the dependence of the actual false alarm rate on the uncertainties of the ETA is also not simple: since the larger timing uncertainties dominate the time window used for the coincidence search (see Section~\ref{coincidences}), the role of the detectors showing the worst timing performance is enhanced. 

\subsection{Event time series statistics}
\label{sec:times}

In a time coincidence search, the statistics of the estimated time of arrival of the events plays a fundamental role. In case the event times are random, a Poisson point process would fit the data. Actually, the data do not reproduce a homogeneous (i.e. stationary) point process, due to the changing performance of the detector and to the statistics of outliers. For our purposes, it is enough to check timescales below $\sim 1$~hour, because they are critical for the method implemented to estimate the noise background (see Section~\ref{sec:background}).

\begin{figure*}[!htb]
\centering
\includegraphics[width=175mm]{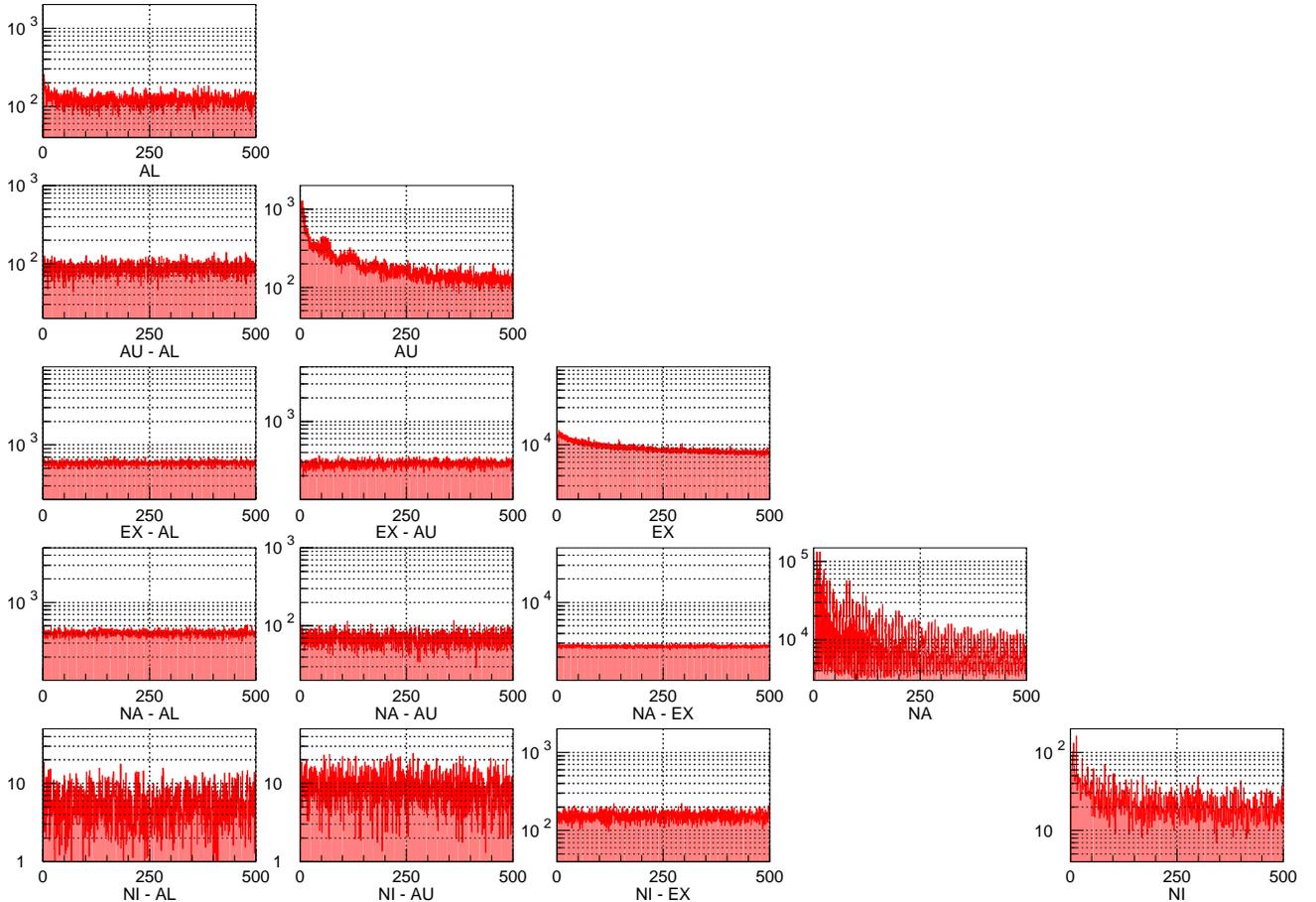}
\caption{\label{fig:correlograms}Correlogram matrix for the estimated time of arrival of all exchanged events in the years 1997-2000; all plots are histograms of the time lags between events in units of counts/sec (\textit{ordinate}) vs seconds (\textit{abscissa}). All five auto-correlation histograms of the single detectors (\textit{diagonal plots}) show correlation at short timescales, and in three cases these structures dominate by far over the uniform distribution expected in case of a Poisson process. Despite this, the cross-correlograms (\textit{off-diagonal plots}) show no sign of residual correlation. The NA-NI cross-correlogram is empty because these two detectors had no overlap in their operative time.}
\end{figure*}

In order to investigate the ETAs statistics of IGEC exchanged data we calculate their correlation histograms (or \textit{correlograms}) as shown in FIG.~\ref{fig:correlograms}. Correlograms are histograms of the time lags between ETA's. For a Poisson process one expects the histogram to be flat --i.e. without preferred time delays between events (see appendix \ref{app:correlograms} for details). It turns out that the ETAs show relevant auto-correlation at small timescales, down to a few seconds. A clustering of the event time series for all the five IGEC detectors is evident. The clustering disappears as soon as one looks to the cross-correlation properties. This is very relevant for the next phase (i.e. coincidence search), because if this were not the case, the output of the empirical method we use to estimate the background would be biased (see Section~\ref{sec:background}).

\section{Multiple detector analysis}
\label{sec:multidet}

In this Section we present the methods implemented by IGEC to analyze the exchanged data. The analysis is based on a time coincidence search among gw candidates, or events, of different detectors (see~\ref{coincidences}). The time coincidence window is varied to get the desired maximum probability of false dismissal. There is also a test of the compatibility of the signal amplitude, as estimated by the different detectors (see~\ref{sec:amplitudecheck}). Prior to the coincidence search, we apply a data selection procedure which limits the search to burst gws exceeding a specified search threshold (see~\ref{sec:preprocessing}). A directional search strategy is implemented as well. The IGEC analysis is then performed as a function of the search threshold value and of the gw direction. The main advantages of this procedure are a reduction in the false alarm rate and the control of the maximum probability of false dismissal of any burst gw exceeding the search threshold. The accidental coincidence background has been estimated by applying the same analysis procedures on many data sets obtained by shifting the time of the real data (see~\ref{sec:background}). The first relevant result of this analysis is that the number of estimated accidental coincidences turns out to be a Poisson random variable. Finally, in ~\ref{sec:confidence} we describe the statistical method used to set the confidence intervals on the number of gw bursts detected by the observatory. This method is unified and ensures the desired coverage of the resulting confidence intervals.

\subsection{Data selection}
\label{sec:preprocessing}

Before searching time coincidences, we apply a data selection procedure which limits the search to burst gws exceeding a specified amplitude. 

The first step is to specify an absolute threshold $H_t$, or \emph{search threshold}, for the gw amplitude estimates.  This threshold is common to all detectors and sets the lower bound on the amplitude of the target gw population. The following multiple detector analysis is then repeated systematically for different threshold values.

The second step is to exclude from the observation time of each detector all time periods where the exchange threshold is \emph{above} the chosen search threshold. The motivation is to limit the false dismissal probability of any burst gw of amplitude greater than the selected threshold. With this selection, the detection efficiency for burst gws exceeding $H_t$ is at least 0.5 in any detector.

The third step is to exclude candidate events that are \emph{below} the search threshold. This gives a significant reduction of the false alarms, while preserving the same minimum detection efficiency of the previous step. In fact, the exclusion of the lower amplitude events cuts down both the rates and the time uncertainties of the events of each detector (see FIG.~\ref{fig:meanrate} and FIG.~\ref{fig:meanMT2}).

\begin{figure}
\centering
\includegraphics[width=85mm]{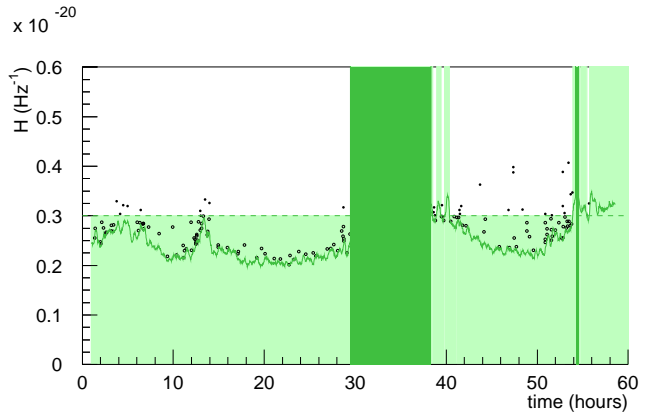}
\caption{\label{fig:ampselection}An example of the selection on AURIGA data for a search not optimized for a specific gw direction. The continuous gray line shows the amplitude of the exchange threshold, the dots represent the exchanged events vs. time. In the dark gray period no data was exchanged. The light gray shaded area shows the amplitude-time regions excluded by the data selection at a search threshold $H_t = 3 \cdot 10^{-21}Hz^{-1}$. The observation times with exchange threshold $> 3 \cdot 10^{-21}Hz^{-1}$  are now excluded as well as the events with amplitude $< 3 \cdot 10^{-21}Hz^{-1}$.}
\end{figure}

This procedure differs from what have been previously done in the field \footnote{In some previous analyses~\cite{Amaldi89,ALEX99}, a systematic study has been made as a function of the amplitude of the events but not of the sensitivity of the detectors so to ensure a minimum detection efficiency. In the others~\cite{Coinc1,Coinc2,Coinc4}, the observation time has been selected according to some fixed threshold on the detector noise and no common thresholding on the amplitude of events have been performed.}.
The data selection is illustrated in FIG.~\ref{fig:ampselection} with a sample of AURIGA data. In general, a higher search threshold $H_t$ allows new portions of the observation time to be considered, thus increasing the effective observation time. Moreover, the rate of the selected events will strongly depend on the value of the detector exchange threshold with respect to $H_t$. In particular, higher rates of events are favored whenever the exchange threshold approaches and crosses the search threshold.


The described data selection is suitable for a blind search over the sky, irrespective of the source location. In fact, a burst gw is seen at each detector with the same amplitude, given that the antenna patterns are almost coherent. 

\begin{figure}
\centering
\includegraphics[width=85mm]{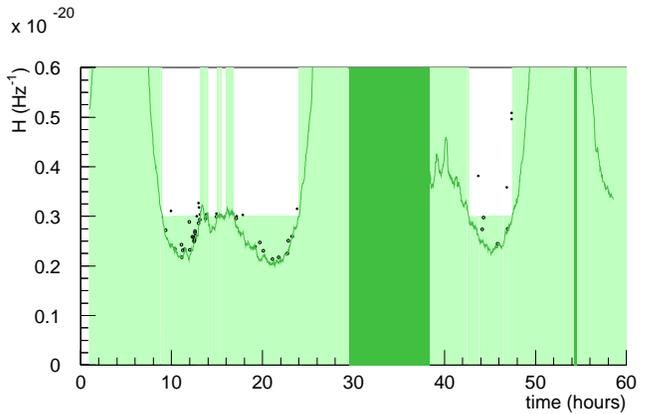}
\caption{\label{fig:ampselection_GC} The same data of FIG.~\ref{fig:ampselection} but now specialized to a  search for gravitational wave bursts from the Galactic Center. The exchange threshold (\textit{gray line}) and events (\textit{dots}) have been divided by the amplitude directional sensitivity of the detector to the Galactic Center. The resulting amplitudes are in terms of a burst gw from the Galactic Center. The light gray shaded area shows the amplitude-time regions excluded by the data selection at a search threshold $H_t = 3 \cdot 10^{-21}Hz^{-1}$.}
\end{figure}

It is also possible to implement a directional search strategy to optimize the search for a specific gw direction. This has been accomplished by modulating the exchanged data with the directional sensitivity of the detectors. Specifically, all exchanged amplitudes (event amplitudes, exchange thresholds, etc.) are divided by the time-dependent angular attenuation factor for the specific direction in the sky (see FIG.~\ref{fig:pattern}). In this way, all amplitudes are given in terms of a burst gw propagating from the selected direction. FIG.~\ref{fig:ampselection_GC} shows the search for burst gws from the Galactic Center direction on the same data set of FIG.~\ref{fig:ampselection}. Then, the rest of the selection procedure above is applied. The effective observation time is reduced with respect to that obtained for a blind search over the sky at the same threshold $H_t$ because the periods when the detector is not favourably aligned with the source are removed. Additionally, the set of selected events is generally different.

As a result, the background noise is reduced while the detection efficiency for the selected direction is preserved. It is worth noticing that the angular selectivity of any directional search is quite poor, due to the broadness of the directional sensitivities of the detectors. For instance, when the detector is optimally oriented with respect to the chosen direction, any source within $\pm 20deg$ from it is seen with at most 11\% attenuation.

The modulation by the directional sensitivity correlates the amplitudes of events and exchange thresholds among different detectors in a much more significant way than any other observed daily effect. This modulation produces new cuts to the observation time and related clusters of events which are almost synchronized in different detectors. As a consequence, the probability of coincidences is enhanced at the edges of the time spans of common observation. The effects of this when estimating the rate of accidental coincidences is discussed in Sec.~\ref{sec:background}.

We note that the false dismissal contributed by this data selection is at most 50\% per each detector for burst gws of amplitude exceeding the chosen search threshold $H_t$. The overall false dismissal of this multiple detector analysis will be further increased by the next two steps, namely the time coincidence search and the amplitude consistency check (see the following Sections). However, the overall false dismissal for a burst gw above $H_t$ is under control at least in a conservative sense. The optimal range of $H_t$ values that would enhance the chances of a gw detection will depend also on the (unknown) amplitude distribution and rate of the burst gws.


\subsection{Time coincidence search}
\label{coincidences}
Two events from different detectors are defined to be in coincidence if their estimated times of arrival $t_i $ and $t_j$ are compatible within their variances $\sigma _{t_i }^2 $ and $\sigma _{t_j }^2 $:
\begin{equation}
\label{eq:coincidence}
\left| {t_i  - t_j } \right| \le \Delta t_{ij}  \equiv \left( {k\sqrt {\sigma _{t_i }^2  + \sigma _{t_j }^2 }  + \Delta t_{MAX} } \right)
\end{equation}

where $\Delta t_{ij}$ is the coincidence window and $\Delta t_{MAX}$ is the maximum expected light travel time between the detectors. $\Delta t_{ij}$ is computed according to the desired probability of false dismissal by setting $k$ through a Bienaym\`e-Tchebyscheff inequality. 

Specifically, the probability to miss a coincidence because of the uncertainties in the ETA is
\begin{equation}
FD \equiv P\left\{ {\left| {t_i  - t_j } \right| \ge \Delta t_{ij} } \right\} \nonumber
\end{equation}
which in turn is upper bounded by
\begin{equation}
\label{eq:PT}
FD \le P\left\{ {\left| {t_i  - t_j } \right| \ge k\sqrt {\sigma _{t_i }^2  + \sigma _{t_j }^2 } } \right\} \le \frac{1}{{k^2 }} \equiv P_T
\end{equation}

where $P_T$ is the maximum false dismissal probability chosen for the coincidence search (see Appendix~\ref{app:bienayme}). The resulting coincidence window changes for each couple of events according to their ETA variances. 

This procedure ensures that the required false dismissal probability is met regardless of the distribution of the estimated arrival time uncertainties. In fact, it is not possible in general to approximate the ETA statistics as Gaussian, due to either the intrinsic narrow bandwidth of the detectors or their limited time resolution. For instance, in the AURIGA data the time accuracy has been 1 ms and the time uncertainty distribution is multi-modal \cite{TimingNPB} at the $SNR$ of interest. The ETA standard deviation is then dependent on the $SNR$ of the event. We note that $\sigma_t  \gg \Delta t_{MAX} $ is the standard condition in these IGEC data. For comparison, the previous IGEC preliminary analysis \cite{IGECPRL} has been performed with a fixed value for the coincidence time window and therefore only a vague indication of the false dismissal was possible. 

The implemented coincidence search allows one event to be in coincidence with more than one event in the other detector. A coincidence in more than two detectors has to satisfy eq.~\ref{eq:coincidence} for all combinations of detector pairs, and the resulting conservative false dismissal has to take into account the number of such required conditions. The same time coincidence search algorithm has been applied both to actually search for burst gws and to estimate the corresponding number of accidentals (see Section~\ref{sec:background}).

The choice of the conservative false dismissal $P_T$ due to the coincidence search can be optimized (see Appendix~\ref{app:bienayme}). It turns out that $P_T$ should be set between 30\% and 5\% to achieve a satisfactory balance between false alarm and false dismissal probabilities in a two-fold coincidence search.

\subsection{Amplitude consistency check}
\label{sec:amplitudecheck}
The estimated amplitude of a burst gw is affected by the systematic and statistical uncertainties of the detector. Hence, once a set of events are found to be in coincidence, one can test if the differences of their estimated amplitudes are consistent with zero within a chosen confidence level. The goal is to lower the false alarms by removing the accidental coincidences whose amplitudes are not consistent. The test we implemented is similar to the condition described above to define a time coincidence \footnote {Different amplitude consistency tests has been applied to coincidence searches with other data sets. The procedures and the results are reported in ref. \cite{amplituderatio,Coinc2,Coinc4}.}. It takes into account both the variance $\sigma _{A_i }^2 $ and the 4th central moment $\mu _{A_i }^{(4)}$ of the estimated amplitude of the events $A_i $, which are included in the current IGEC exchange protocol.

Two events from different detectors have consistent amplitudes if
\begin{equation}
\left| {A_i  - A_j } \right| \le {\Delta A_{ij} + \Delta A_i  + \Delta A_j }
\end{equation}
where
\begin{equation}
\Delta A_{ij}\equiv \min \left\{ {\sqrt {\frac{{\sigma _{A_i }^2  + \sigma _{A_j}^2 }}{{P_A }}} ,\nonumber
\sqrt[4]{{\frac{{\mu _{A_i }^{(4)}  + \mu _{A_j}^{(4)}  + 6  \sigma _{A_i }^2   \sigma _{A_j }^2 }}{{P_A }}}}} \right\}
\end{equation}

and $\Delta A_i$ are the systematic amplitude calibration errors of the detectors. $P_A$ is the required conservative false dismissal of the test for a coincidence which corresponds to a burst gw. The two alternative terms in curly brackets come from the Bienaym\`e inequality of 2nd and 4th order respectively applied to the random variable $A_i-A_j$. The more stringent of them is chosen time by time (see Appendix~\ref{app:bienayme}).

The general Bienaym\`e inequality is used because the amplitude noise distributions of the detectors were not Gaussian nor modeled for a significant fraction of the observation time. 
This test is conservative and provides a less stringent removal of false alarms at low $SNR$ amplitudes with respect to tests based on some modeled statistics. Instead, at high $SNR$ amplitudes the systematic calibration errors, 10\% for all detectors, dominate over $\Delta A_{ij}$.


The efficiency of the false alarm rejection of this amplitude consistency test was found to be strongly dependent on the search threshold. The false alarms were significantly reduced only at high thresholds,$H_t \geq 1\cdot10^{-20}Hz^{-1}$. This is due to two concurrent facts. First of all, the implemented data selection forces most of the  events to have similar amplitudes, since they are constrained from below by the imposed search threshold and from above by the steep slope of the amplitude distribution (Fig.~\ref{fig:amplitude}). Therefore, the data preprocessing itself provides an implicit rejection of most of the events in coincidence which show non--consistent amplitudes. Second, the efficiency of the test is greater at high $SNR$ amplitudes, because there the amplitude differences of accidental events can be larger in terms of standard deviations. 

As a result, the test turned out to be convenient only at high search thresholds and with values of conservative false dismissal $P_A < 30\%$. On the contrary, at low thresholds the implementation of the test is disadvantageous because the increase in the false dismissal is not balanced by a sufficient false alarm reduction. For the results described in this paper, the application of this amplitude consistency test did not add new significant information. 

\subsection{Background estimation}
\label{sec:background}
In order to assess the statistical significance of the number of detected coincidences, a reliable estimate of the background --i.e. the number of coincidences that have been found by chance-- is needed. The ideal approach would be to obtain new independent data sets from a population with the same statistics but with all gw sources switched off, to repeat the search procedure, and then to compare the statistics of the found coincidences with those obtained from the original sample.

If the ergodic hypothesis applies to our data, a good method to create an independent data sample is to perform a relative translation of the time coordinate of the data exchanged by different detectors. This operation preserves the statistics of the single event list (average number of events, instantaneous rate fluctuations, auto-correlation of event times, ...). The coincidence counts found on time shifted data sets are independent as long as the applied time delays are longer than the maximum time window used in the coincidence search. It is reasonable to assume that the gw events are a negligible fraction of the \textit{total} number of events at each detector and therefore the gw events will not affect the coincidences within the shifted data sets. This empirical method has been widely used in the field (see for instance, \cite{Amaldi89,ALEX99,Coinc1,Coinc2,IGECPRL}). We remark that this framework allows us to only estimate the level of the coincidence background, but not to distinguish between coincidences due to gw and those due to other common sources. 

There are a number of technical subtleties to address in order to give a complete description of the way this method has been practically implemented in this analysis. One issue regards the common observation time of each time shifted configuration, i.e. the collective length of the time spans after the data selection phase. In fact, in case of significant changes of the shifted observation time one should consider the \textit{rates} rather than the counts of accidental coincidences, but then, the statistics of this random variable is no longer Poissonian.

Another crucial issue is to check the statistics of the background, especially in IGEC data since the statistics of the event time series of each detector shows evidence for auto-correlation at small timescales (see FIG.~\ref{fig:correlograms}). We require that the statistics of the coincidence background estimates be Poissonian and stationary for any applied time shift, as is expected if the coincidence time series can be modeled by a Poisson point process. In fact, this is the model we use in Section ~\ref{sec:confidence} to estimate the statistical significance of the found coincidences with respect to the background. Note that the instantaneous coincidence rate may be also time-varying (i.e. non-homogeneous), and still the coincidence counts in a fixed time span of a shifted configuration would be a sample of a Poisson random variable. Independence (i.e. randomness) of successive coincidence times is the key to guarantee this result.
This requirement is met when the event lists are not cross-correlated within timescales up to the maximum applied time shift.

When performing a search optimized to a specific direction (Section~\ref{sec:preprocessing}), we pointed out the additional problem related to the appearance of clusters of events at times correlated in different detectors. In this case the key for a successful estimation of the background has been to time shift the data \textit{before} applying the amplitude modulation.

\begin{figure}
\centering
\includegraphics[width=85mm]{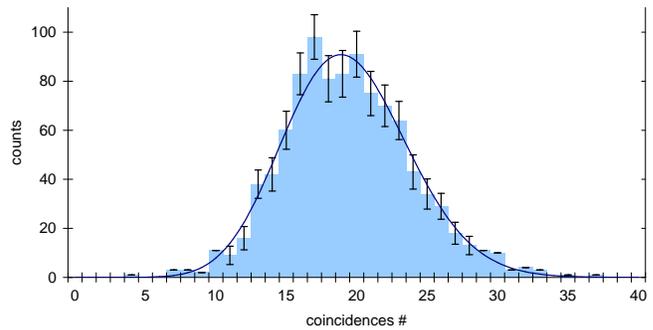}
\caption{\label{fig:poissonfit}Sample of the histogram of the background coincidences obtained by the time shift method with a superimposed Poisson fit. The plot shows the shifted coincidences between EXPLORER and NAUTILUS searched with $P_T=5\%$, $P_A=0$ and search threshold  of $5.62\cdot 10^{-21}Hz^{-1}$ without optimizing for a gw direction. The one tail $\chi^2$ probability of the sample is 0.7. The $\chi^2$ test is performed only on the histogram bins with at least 10 counts. The maximum time shift reached $\pm 5000 s$ in $10 s$ steps.} 
\end{figure}

\begin{figure}
\centering
\includegraphics[width=85mm]{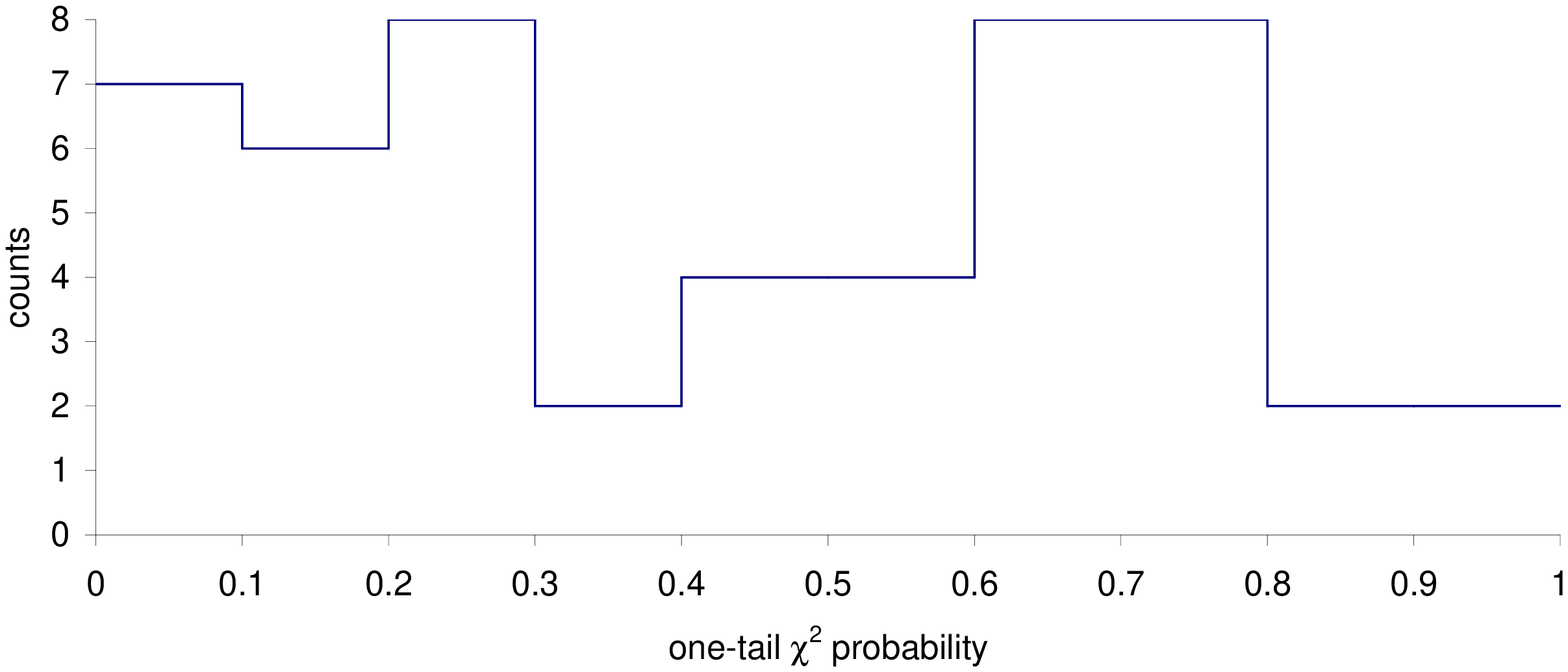}
\caption{\label{fig:poissonchi2}Histogram of the significance level of the goodness-of-the-fit test for the Poisson model of the background estimates (one tail $\chi^2$ probabilities). The $\chi^2$ test has been performed on all the configurations of the observatory which ensured at least one degree of freedom (see FIG.~\ref{fig:poissonfit}). The plotted data include configurations of different detectors with different search threshold values and same conservative false dismissal of the coincidence search ($P_T=5\%$, $P_A=0$), without optimizing for a gw direction. The maximum time shift reached 5000s in 10s steps. The histogram is well described by a uniform distribution, so we can conclude that the $\chi^2$ tests give results consistent with a general agreement of the background statistics to the Poisson model.}
\end{figure}

In this analysis, the number of tested time-shifted configurations for a pair of detectors is $N_s\sim 1000$. This number is limited by two requirements. The first is to keep the shift step larger than the longer time window used for coincidence search. On the other hand, the maximum delay has to be small enough to keep stationary the random variable number of accidental coincidences. The maximum time shift has been limited to $\sim 5 \cdot 10^3$ seconds and the shift steps has been between 5 and 15 seconds.

The easiest way to check the Poisson statistics and the independence of the background samples is to create an histogram of the number of coincidences found at each time shift and fit it to a Poisson density function. FIG.~\ref{fig:poissonfit} shows the agreement of the background statistics to the model for a sample configuration. To test the goodness of the Poisson fits we applied chi-square tests whenever the histograms of the background counts were sufficiently populated. The probability of getting a chi-square greater than the observed value has to be a sample of a uniform density between 0 and 1 if the performed fit is good. This has been actually confirmed, as shown by the histogram of the resulting one-tail chi-square probabilities shown in FIG.\ref{fig:poissonchi2}; therefore, we can conclude that no deviations of the background estimates from the expected Poisson statistics are observed. 

We remark that this is a quite relevant result, since in our case the single event lists showed evident auto-correlation at short timescales (see FIG.~\ref{fig:correlograms}) and since the gw search from a specific direction brings in a significant correlation of the selected event rates in different detectors, as described in Section~\ref{sec:preprocessing}. As a consequence, we have been able to use all the available data to set confidence intervals on the detected gw with the procedure described in the following Section.

\subsection{Setting confidence intervals on detected GW}
\label{sec:confidence}
The found number of coincidences and the expected background are compared under the hypothesis of a superimposed homogeneous Poisson rate of detected gw signals. The results are then expressed as confidence intervals containing the true rate of detected gw with a given probability, i.e. assuring a given \textit{coverage}. The procedure we adopt to set confidence intervals follows the track of previously reported \textit{unified} methods \cite{FeldmanCousins1998,RoeWoodroofe2000}. The results will be interpreted either as upper limits or as evidence for detection in case the confidence intervals include or not the null result.

We assume a Poisson model for the coincidence background, whose expected number $\mathbf{\bar N}_b$ is estimated as described in the previous subsection. Having observed a number of actual coincidences $N_c$, the likelihood function of the average number of detected gw, $N_\Lambda$, is: 
\begin{eqnarray}
{\ell (N_\Lambda  ;N_c, \mathbf{\bar N}_b )} \equiv \left\{ {
\begin{array}{*{20}c}
 {  P_{N_c } (\mathbf{\bar N}_b + N_\Lambda ) }&{ \textnormal{if}\quad N_\Lambda \ge 0  }\\
   {0} & {\textnormal{if}\quad N_\Lambda < 0}
\end{array} }  \right.
\end{eqnarray}

where

\begin{equation}
P_{N_c } (\mathbf{\bar N}_b  + N_\Lambda  ) \equiv \frac{1}{{N_c!}}\left( {
\mathbf{\bar N}_b  + N_\Lambda  } \right)^{N_c } e^{ - \left( {\mathbf{\bar N}_
b  + N_\Lambda  } \right)}
\end{equation}

Then the confidence interval is defined by integrating the likelihood over the smaller domain $\left[ { N_{\inf } \div N_{\sup } }\right]$ which ensures that the integral amounts to a specified value I.

\begin{equation}
I = \left[ {\int_0 ^\infty  {\ell (N)dN} } \right]^{ - 1} \int_{N_{\inf } }^{N_{\sup } } {\ell (N)dN}
\end{equation} 

The intervals defined as such have nice properties \cite{RoeWoodroofe2000}. First, they are naturally bound to the physical domain. Moreover, they include the most likely estimate of the number of detected gw, which is zero when the expected background exceeds the found number of coincidences. Such most likely confidence intervals are also the \textit{most credible} in the Bayesian framework assuming a uniform prior for $N_\Lambda\geq 0$. However, when the value for the likelihood integral I is properly chosen, the resulting interval has also a well-defined minimum frequentist coverage. From numerical computations it turns out that $I=0.94$ and $I=0.97$ guarantee the coverage to be at least 0.90 and 0.95 respectively. By taking into account the effective observation time, the confidence intervals can be expressed in terms of the Poisson rate of detected gws.

The overcoverage ensured by this unified method is quite significant for true gw rates much less than the background rate. In particular, in case the true value is exactly zero (null hypothesis) the complement of the coverage to unity can be interpreted as the false detection probability, and it is lower than $\sim 4.5\%$ and $\sim 2.3\%$ when the conservative coverage is respectively 0.90 and 0.95.

\section{Results}
\label{sec:results}

The IGEC observations have been analyzed both by performing a blind search over the sky (without selecting a specific gw direction) and by optimizing the search for burst gws from the Galactic Center direction. In the first case the results refer to the amplitude component of the burst gw along the detector axes. In the second case the results are given in terms of the amplitude of a burst gw from the Galactic Center. In both cases, only the polarization component along the bar axis is considered.

We define the operating time of a particular \textit{configuration} of detectors to be the subset of the network operation periods when only the detectors of this configuration are simultaneously operative. The main advantage of this procedure is that the results from different configurations are then automatically {\it independent}, since they refer to disjoint observation times. We remark that the operating times of the configurations depend on the search threshold (see Section~\ref{sec:preprocessing}). Within IGEC, 18 different configurations of detectors have been operating during 1997-2000: 9 pairs, 7 triples and 2 four-fold configurations.  The multi detector data analysis has been performed separately for each configuration as a function of the search threshold $H_t$ in the range $2-50 \cdot 10^{-21} Hz^{-1}$.  

In order to synthesize the overall result of the observatory at each threshold value, we sum the observation times, the coincidence counts and the backgrounds over all the configurations at the same $H_t$. A confidence interval of the whole network is then re-computed accordingly for each $H_t$. 

We report a synthesis of the results (number of coincidences, background, observation time, confidence interval on detected gws) for each configuration of detectors and investigated threshold value in Appendix~\ref{app:tables}. The results depend also on the choice of the parameters of the analysis, namely the false dismissal on the time coincidence search, the false dismissal on amplitude consistency of events in coincidence, the selection of a direction in the sky and the required conservative probability, or \textit{coverage}, of the confidence intervals. 

The results obtained at each threshold value are cumulative, i.e. they apply to detected burst gw whose amplitudes are $\geq H_t$. This is a consequence of the data selection described in Section~\ref{sec:preprocessing}. In particular, an upper limit set at some threshold $H_t$ is valid with the same conservative coverage for any higher threshold. 

\subsection{No statistical evidence for detected GW}
\label{sec:noevidence}
The overall results over the entire time span 1997--2000 are well in agreement with the estimated background. In fact, the resulting confidence intervals on the number of detected gw signals include the null result in almost all the many trials performed (see Appendix~\ref{app:tables}). Only a few 2-fold configurations give  gw detections for some specific values of the parameters of the analysis. We will show, however, that the relative frequency of these cases is well accounted for by the probability of false alarm, i.e. of getting by chance a detection in case no gw were present in the data. 

The probability that a confidence interval may fail to include the null result in case no gw are present in the data has been numerically computed by simulating the procedure to set confidence intervals (see Section~\ref{sec:confidence}). Due to the overcoverage of these confidence intervals, this probability is always smaller than the maximum false dismissal associated to the interval, i.e. $1-coverage$, especially at low background levels. Specifically, when the accidental coincidences are $\mathbf{\bar N}_b\ge 1$, the expected number of false alarms oscillates around $\sim 3\%$ and $\sim 1.5\%$, respectively for maximum false dismissal values of 10\% and 5\%. This means that false alarms of the order of one every 30 (resp. 70) independent trials are the rule in case of $\mathbf{\bar N}_b\ge 1$. Instead, in the limit of low background, i.e. $\mathbf{\bar N}_b<0.01$, the false alarm probability turns out to be $\mathbf{\bar N}_b$ regardless of the required coverage. Therefore no false alarms are expected for the configurations showing low enough background levels. These predictions on false alarms obtained by numerical simulations have also been confirmed by an independent empirical method (see Appendix~\ref{app:tables}).

No statistical evidence for detected gw has been found. In fact, on one hand the null result is always included in the confidence intervals corresponding to low background levels, such as those related to triple and four-fold configurations of detectors. On the other side, the total number of detections is consistent with the total number of expected false alarms. For the many trials reported in Appendix~\ref{app:tables}, the expected \{found\} total numbers of false alarms for the Galactic Center search are 3.5 \{2\} and 1.7 \{0\} for 0.9 and 0.95 coverage respectively. For the blind search over the sky these numbers are 2.5 \{4\} and 1.4 \{2\}. 

With at least three IGEC detectors in simultaneous operation the false alarms are extremely rare even at search thresholds $H_t$ close to the exchange thresholds of the single detectors. So, even after many years of observation time, such configurations would allow to easily identify any detected gw. Instead, in the case of two-fold coincidence search, the false alarms populate the achieved observation time up to high search thresholds, $H_t\le 10^{-20} Hz^{-1}$. 

\subsection{Upper limit on the rate of detected GW}

\begin{figure}
\centering
\includegraphics[width=85mm]{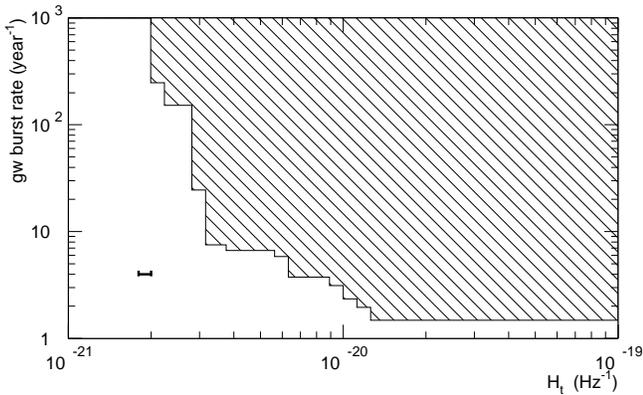}
\caption{\label{fig:rate_noGC}Upper limit for the rate of the burst gws detected by the observatory as a function of the amplitude search threshold $H_t$. No directional search has been applied (i.e. blind search over the sky). The Fourier amplitude in abscissa refers to the component of a burst gw along the axes of the detectors. The dashed region above the continuous line is excluded with at least 95\% probability. The uncertainties on the estimated background affect neglibly the plotted curve. The time window for the coincidence search has been selected to limit the related false dismissal probability to at most 5\% and no test on the amplitude consistency between events has been applied. The maximum amplitude systematic error related to the calibrations of the detectors is shown as the error bar parallel to the abscissa.}
\end{figure}

\begin{figure}
\centering
\includegraphics[width=85mm]{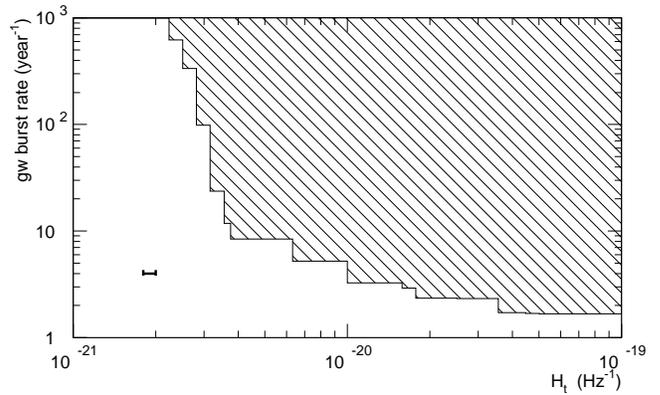}
\caption{\label{fig:rate_GC}Same as FIG.~\ref{fig:rate_noGC} but specialized for the rate of burst gws from the Galactic Center direction. The Fourier amplitude in abscissa refers to the amplitude of a burst gw from the Galactic Center direction with optimal polarization with respect to the detectors.}
\end{figure}

The resulting IGEC observations can be synthesized by an upper limit on the rate of detected burst gw, modeled as a Poisson point process with constant rate. This upper limit as a function of the signal search threshold is given in FIG.~\ref{fig:rate_noGC} and ~\ref{fig:rate_GC} for a search not optimized for any specific direction and optimized for the Galactic Center direction respectively. These upper limits are based on the confidence intervals of the network, computed from the sums of the data (observation times, coincidence counts and background) of all configurations per each threshold value. The plotted upper bounds have a probability of at least 95\% to be greater than the actual gw rate value, so that the dashed region is excluded with the same confidence. The lower limits of the confidence intervals are all at null gw rate but one per each type of search (Galactic Center and blind). These detections can however be explained as expected false alarms (see previous Section and Appendix~\ref{app:tables}) and do not affect the upper limits shown in the Figures.

The upper bounds set by the network show in a few cases higher gw rates at higher thresholds than at lower thresholds (see Appendix~\ref{app:tables}). This happens for the same reason why Fig.~\ref{fig:meanrate} and Fig.~\ref{fig:meanMT2} may show increasing event rates and timing errors as $H_t$ increases. In these cases, we chose to consider the most stringent upper bound value, on the basis of the fact that an upper limit computed at some threshold is also valid for any higher value of the threshold -- as already remarked. This choice introduces a marginal bias on the stated coverage close to the plotted upper bounds.

The results of the blind search, FIG.~\ref{fig:rate_noGC}, show a flat upper limit on the gw rate at high search thresholds, $H_t > 10^{-20} Hz^{-1}$, where no coincidences have been found. This rate is determined by the \textit{total} observation time of the network at those search thresholds, $T_{obs}$ and takes into account the conservative false dismissal of the time coincidence search. Specifically, the rate is given by $ F\ /(T_{obs}\ CL_{c})$, where $CL_{c}$ is the confidence level of the time coincidence search procedure, $CL_{c}=0.95$ for the data plotted, and the factor $F$ depends on the required coverage, $F=3.6$ for 0.95 probability. The confidence intervals on the rate widen at intermediate search thresholds mainly because of the presence of accidental coincidences. At $H_t \leq 3\cdot 10^{-21}$ the upper limit sharply increases because of the corresponding cut off on the observation time. The results referred to the Galactic Center direction, FIG.~\ref{fig:rate_GC}, show a similar behavior, smoothed by the effect of the modulation of the directional sensitivity of the detectors. The relevant data are also tabulated in Appendix C.

\subsection{Final remarks}
The reported results refer to the gw detected by the observatory, that is to say to the gw present as coincidences in the IGEC exchanged data during the 1997--2000 observation time. In order to extend these results to the flux of gw crossing the earth, one should take into account the actual efficiency of detection, which also depends on the specific model of the gw source, in particular on the gw amplitude distribution and rate. However, this goes beyond the scope of the present paper. Here we set conservative detection efficiencies of the data analysis procedures. 

The main progresses over the first IGEC analysis~\cite{IGECPRL} rely on the optimization with respect to amplitude and  direction of the gw, on the balance between false dismissal probability and false alarm background, on the assessment of the statistical coverage of the upper limits and on the more extended observation time. The recent findings from the NAUTILUS and EXPLORER 2001 data~\cite{Coinc4} cannot be compared with the results of the analysis presented in this paper, as they relate to detectors runs with higher sensitivity.

We remind once more the reader that we are giving to the confidence intervals a frequentist statistical interpretation: they are determined to the best of our knowledge in order to include the actual value of the detected gw rate with a relative frequency given \textit{on average} by the chosen coverage. The method and the likelyhood from which the confidence intervals are computed is described in Section~\ref{sec:confidence}. An alternative approach based on the Bayesian framework can be followed for the data analysis. Section~\ref{sec:confidence} gives all the necessary information, in particular the most credible intervals assuming a uniform prior. These credible intervals would result similar to the ones we presented, but the value for their confidence and its interpretation would be different. For instance, the presented confidence intervals with 90\% coverage would also be the most credible intervals with degree of belief of 94\%, but different priors can be used as well.

IGEC amplitude data are in terms of the Fourier component $H$ of the strain amplitude $h$ of the gw burst signal. The relation between $h$ and $H$ depend on the specific model of signal shape. For instance, for a signal consisting of one sinusoidal cycle of 1 ms period $h\simeq (2 \cdot 10^3Hz)\cdot H$. For such a source located at the Galactic Center and emitting isotropically, the estimated mass converted in gw would be $\sim 0.01 M_{\odot}\cdot (H /1.5\cdot 10^{-21}s)^2$.

The exchanged IGEC data do not allow the measurement of the light travel time of gw signals among the detector sites. In particular, past IGEC observations cannot resolve the gw direction. This capability will be achieved when upgraded detectors will demonstrate wider sensitivity bandwidths, much greater than $10 Hz$, and will implement signal acquisition and filtering ensuring sub-millisecond time resolution\cite{TimingNPB}. In fact, under these conditions, we expect that the overall uncertainties in estimated arrival time of the events will be much smaller than the light travel time between detector sites. Progress in this respect has been recently achieved\cite{EXPLORER2002} and are expected for the next runs of bar detectors\cite{AURIGAdue, EXPLORER2002}.

\begin{acknowledgments}
We wish to acknowledge the assistance of G.Maron and M.Biasotto, who set up the linux pc farm used for the most intensive computations at INFN -- Laboratori Nazionali di Legnaro. We are indebted to A.Colombo, M.Lollo and G.Soranzo for their skilled technical help during the set up and operation of the gw detector AURIGA. This work has been partially funded by a grant COFIN 2000 from the Italian MIUR. The work at LSU was funded by the National Science Foundation and by LSU. NIOBE was supported by the Australian Research Council.
\end{acknowledgments}

\appendix

\section{Time series and correlograms}
\label{app:correlograms}

Given two uniform and uncorrelated random point processes $\left\{ {t_1^i } \right\}_{i \in \mathbb{N}}$ and $\left\{ {t_2^i } \right\}_{i \in \mathbb{N}}$, the set of ordered time differences $\left\{ {\Delta ^{ij}  = t_1^i  - t_2^j } \right\}_{t_1^i  > t_2^j }$ between all possible pairs is described by a uniform distribution. This can be displayed by projecting all values in $\left\{ {\Delta ^{ij} } \right\}$ into an histogram. The bin content is the \textit{total} number of points of the original process $\left\{ {t_2^i } \right\}_{i \in \mathbb{N}}$ that fall within a range (equal to the bin size) at a specified  time lag from each of the points in the series $\left\{ {t_1^i } \right\}_{i \in \mathbb{N}}$. As the two series are independent, this number is a Poisson variable, and its average value is bilinear in the number of events in $\left\{ {t_1^i } \right\}_{i \in \mathbb{N}}$ and $\left\{ {t_2^i } \right\}_{i \in \mathbb{N}}$. We call the histogram of generalized delays between points from different series a \textit{cross-correlogram}, and a \textit{self-correlogram} if $t_1^i  = t_2^i$.

This representation has several advantages over a straightforward simple histogram of the time delays between \textit{successive} points when it is applied to a Poisson point process. A simple time delay histogram describes the first order statistics and, for a perfectly homogeneous Poisson point process, would be fit by an exponential density function. However, this no longer holds true if the rate of the point process varies with time. In addition, any phase correlation at time lags longer than the average time separation is smeared out and not easily identified. A self--correlogram, on the other hand, retains much more information about the auto-correlation of the time series, and is quite insensitive to fluctuations in the rate of the point process that occur on timescales longer than the range of the correlogram.

For instance, consider a non-homogeneous Poisson point process, with a rate $\lambda _1 $ for the first half of the time, and a rate $\lambda _2 $ for the other half. Its first-order delay histogram is the sum of two different exponential distributions, while the self correlogram is still flat (being the sum of two flat distributions). If on top of the random process we add a periodic series with constant rate $\lambda _p  \ll \lambda _1 ,\lambda _2 $, then it would be barely discernible in the first-order histogram, while in the self--correlogram it would appear as a sharp peak at $1/\lambda _p$.

The cross-correlation of two time series is related to the expected background due to accidental coincidences. If a bin of the cross-correlogram has width $dt$ and is centered at time lag $\Delta t$, then the counts inside this bin are proportional to the number of coincidences one would find after a time-shift of one series by $ \pm \Delta t$ and with time window aperture $dt$.

When the cross-correlogram is not flat in a certain range of lags, this means that 
the statistics of the estimated background coincidence counts is not Poisson. Deviations from flatness would suggest a correlation of the rate of the point processes, either because of a common drive acting on more than one detector or because of different local signals but with the same periodic characteristics. A similar feature could also be caused by a border effect due to the time fragmentation of the data series.

\section{Testing coincidences and false dismissal}
\label{app:bienayme}
In order to check if event times and amplitudes in different detectors are consistent with a common gw signal, we use the Bienaym\`e's inequality \cite{Papoulis}. 

We recall that for a random variable $x$ with mean $\eta$, the absolute value of the residual of $x$ with respect to $\eta$ is greater than $\varepsilon$ with probability $P$, given by

\begin{equation}
\label{eq:Bienayme}
P\left\{ {\left| {x - \eta } \right| \ge \varepsilon } \right\} \le \frac{{E\left\{ {\left| {x - \eta } \right|^n } \right\}}}{{\varepsilon ^n }}
\end{equation}

where $E\left\{ {\left| {x - \eta } \right|^n } \right\}$ is the $n$-th absolute central moment of $x$. The Tchebyscheff's inequality is a special case for $n=2$. This inequality holds true for any statistics of $x$, as long as the moments exist. 

When testing that $\eta$ is the mean value, $P$ is the conservative probability of false dismissal, i.e.  of rejecting the hypothesis even though it is true. Concerning our analysis, we invert this inequality to compute $\varepsilon$ given $P$, choosing the most convenient order $n$. $\varepsilon$ can be conveniently expressed in terms of the standard deviation and a non--dimensional multiplier, i.e. $\varepsilon = k \sigma$.

The IGEC exchanged data provide the variance of the estimated arrival time of the events as well as the central moments of $2^{nd}$ and $4^{th}$ order of the amplitude noise distribution. To test the consistency of two values measured with different detectors we apply the Bienaym\`e's inequality to the random variable  $x_i-x_j$, which, in case of events generated by the same gw excitation, has a zero mean, variance $\mu _{ij}^{(2)}\equiv \sigma _{ij}^2 \equiv \sigma _i^2  + \sigma _j^2$ and 4th moment $\mu _{ij}^{(4)}\equiv \mu _i^{(4)}+\mu _j^{(4)}+6 \sigma _i^2 \sigma _j^2$, being $\sigma_i^2$ and $\mu_i^{(4)}$ the $2^{nd}$ and $4^{th}$ order moments of the $i$-th detector. Then, Eq.~\ref{eq:Bienayme} reduces to
$P\left\{ {\left|{x_i  - x_j}\right| \ge k \sigma_{ij} } \right\} \le k^{-n}{\mu_{ij}^{(n)}}/{\sigma_{ij}^n }$.
and the coincidence test $\left|x_i-x_j \right|< k \sigma_{ij}$ is passed with the required maximum false dismissal probability $P$ if we set $k=\sqrt[n]{(\mu_{ij}^{(n)}/\sigma_{ij}^n)/P}$, where the order $n$ is used, which gives the most stringent check (given that moments of order $n>2$ are available).

Regarding the value of the false dismissal $P$, there is an optimal choice to maximize the chances of gw detection. The value $P_T$ used for comparison of event times, eq.~\ref{eq:coincidence} and \ref{eq:PT}, was chosen in order to balance between the conservative probability of detection of a gw coincidence, $1-P_T$, and the related accidentals, whose number is proportional to the time window used for the coincidence search and therefore to $P_T^{-1/2}$. High values for $P_T$ make the coincidence search less efficient, since its efficiency decreases more rapidly than the related background. On the contrary, setting $P_T$ significantly below 5\% has the drawback of gaining too little in terms of detection probability, while increasing significantly the expected false alarms and their fluctuations. The choice $P_T=30\%$ would maximize precisely the ratio of the (conservative) detection efficiency and the corresponding number of accidental coincidences. However, what one should really care is the uncertainty due to  Poisson fluctuations of the background rather than the average background itself. The contribution of these fluctuations when setting a confidence interval depends in a weaker way on $P_T$ (approximately it is given by the square root of the average background). In the end, this would favor lower values for $P_T$, and the optimal choice is expected to be around $5-10\%$. These expectations were confirmed by drawing the final results for different choices of $P_T$.

\section{Analytic tables of coincidence counts and background estimates}
\label{app:tables}

In the following pages the reader can find more detailed information on the many analyses performed in this IGEC search. These trials differ for the directional search (none in Tab.~\ref{tab:noGC}, optimization for the Galactic Center direction in Tab.~\ref{tab:GC}), for the investigated values of the gw amplitude threshold $H_t$, and for the configurations of the network (there where at most 18 disjoint choices of detectors, plus the \textit{total} lines, which synthetize the results of all configurations at the same threshold). For each trial we report the observation time, the corresponding coincidence counts and the background estimates. From these data, and assuming that the bursts can be modelled as a Poisson point process, we compute the confidence intervals on the average number of gw bursts that possibly occurred within that time span (see Section~\ref{sec:confidence}). They are reported in the last two columns of the tables for two different values of the minimum coverage, namely 90\% and 95\%. 

Almost all of the computed confidence intervals cover the null result and therefore, at first glance, there is no strong evidence for gw detection. In order to be quantitative in this conclusion, we should undergo the not-so-easy task of estimating how many \textit{false detections} we should expect in the tables. To do this, one has to consider the exact coverage of the confidence intervals for the specific case of no gw present in the data, rather than the stated conservative coverage, which is the minimum ensured coverage over any possible number of detected gws. Secondly, one must also understand which lines in the tables are really independent and which ones are not.

As already remarked in Section~\ref{sec:noevidence}, the false alarm probability is much smaller than the conservative false dismissal. Considering the results on the whole, the number of detections found are in agreement with the false detections, predicted by summing up the expected false alarm probability of the hundreds of lines in the tables. We must be careful in drawing the conclusions, because the trials we are summing on are not completely independent. The following remarks help to get an idea of the degree of correlation among different lines in the Tables.
\begin{enumerate}
\item
The configurations of detectors at the same search threshold $H_t$ in each Table are independent, since they refer to mutually disjoint time spans. 
\item
Moving toward high $H_t$, as soon as no coincidences are found and the observation time saturates to $100\%$, the results of each configuration do not depend anymore on $H_t$ (for this reason we simplified the Table~\ref{tab:noGC} at  high $H_t$).
\item 
The correlation among outcomes found at different amplitudes can be anything from zero to one. For instance, in Table~\ref{tab:noGC}, stepping from 3.16 to $3.55 \cdot 10^{-20} Hz^{-1}$ the total coincidence count is the same, $N_c=8$, but only 2 are in common.
\item
The correlation between the single configuration outcomes and the related \textit{total} line can also vary, depending on the relative weight of the configurations in the sum. In case the total background counts are mostly due to a particular configuration, then the \textit{total} is quite correlated with it, and therefore the total does not to add much information.
On the other hand, when a few expected total counts are spread in a balanced way among all configurations, then each of them may well end up showing a coincidence even if its specific background is low. These configurations count as many independent trials. Therefore, one should not make the mistake of selecting just those where a coincidence was found, because this would lead to a biased result. Instead, only the \textit{total} line for that threshold should be considered in this case.

\end{enumerate}

For all these reasons, the sparse hints of detection that can be found here and there in the tables have to be criticized from a statistical point of view. In particular, we shall discuss two specific cases in more detail.

The only positive result in a \textit{total} line of Table~\ref{tab:noGC} is at $H_t=3.98\cdot 10^{-21}$ , and seems to confirm the detection suggested by the line \textsc{ex-ni} at the same amplitude. However, when correctly computing the conditioned probability that the \textit{total} line is positive when $N_c(\textsc{ex-ni})=14$ is observed, we find it is as high as 54\% at 95\% coverage. Therefore we shall not claim a detection for this \textit{total} line more than we would because of the single \textsc{ex-ni} result, and we are already aware of its small significance, considering the overall number of expected false detections over the single configurations.  

In Table~\ref{tab:GC} the only detection are at $H_t=3.98\cdot 10^{-21}$: two detections at different configurations for 0.9 coverage and a detection at the \textit{total} line for both 0.9 and 0.95 coverage. In order to compute the the probability of getting a similar result by chance, we devised an empirical method~\cite{baggioGWDAW2002} based on the information available from the time shifts estimates of the accidental coincidences (see Section~\ref{sec:background}). We built a thousand independent tables of results obtained from the samples of time shifted data and we used the resulting statistics of detections as reference for the false alarms. The overall number of detections confirm the predictions reported in Section~\ref{sec:noevidence}). Moreover, the false alarm probability to show detections in at least one \textit{total} line is 0.55 and 0.33 for 0.9 and 0.95 coverage respectively. The probability of getting at least two detections at the same threshold value is 0.15 at 0.9 coverage. All probabilities are therefore well in agreement with no gw detection.

\begin{squeezetable}
\setlength\extrarowheight{2pt}
\setlength{\tabcolsep}{0.5pt}
\setlength{\LTcapwidth}{85mm}
\setcounter{LTchunksize}{50}
\begin{longtable}{|c|c|d|c|>{$}c<{$}|>{$}c<{$}|>{$}c<{$}|}
39.81 &	\textsc{al-au-ex}&	14.0&	0&	<9.1\cdot 10^{-5}&	0\div 2.9&	0\div 3.6 \kill
\caption{\label{tab:noGC}Results of the time coincidence search analysis for all the disjoint configurations of detectors whose observation time has been greater than $1 day$. No directional search has been implemented. The coincidence search has been performed with a conservative false dismissal of 5\%. The estimated average background $\mathbf{\bar N}_b~^{5\%}_{95\%}$ is reported together with its uncertainty, given either as 5\% and 95\% percentiles or as an 90\% upper bound. When followed by the symbol $_*$, the bound has to be read as a conservative upper limit (the actual upper bound is lower). Where the symbol --- appear, there is no available estimate of the average background, though it does not affect the related confidence intervals because in these cases there were no found coincidences. For each search threshold, the bottom \textit{total} line summarizes the results, and is obtained by summing over all configurations the observation time $T_{obs}$, the number of coincidences $N_c$, the background $\mathbf{\bar N}_b~^{5\%}_{95\%}$. In the last two columns, the confidence intervals on the average number of detected gw are reported for 90\% and 95\% conservative coverage. Above the search threshold value $3 \cdot 10^{-20} Hz^{-1}$ the results do not change.}\\
\hline
\hline
\multicolumn{1}{|c|}{H$\cdot 10^{21}$} &\multirow{2}{*}{config}&\multicolumn{1}{c|}{$T_{obs}$}& \multirow{2}{*}{$N_c$}&\multirow{2}{*}{$\mathbf{\bar N}_b~^{5\%}_{95\%}$} & \multicolumn{2}{c|}{confidence interval} \\ 
\multicolumn{1}{|c|}{(Hz$^{-1}$)}& & \multicolumn{1}{c|}{(days)} & &  & 90\%& 95\%\\ \hline
\endfirsthead
\hline
\multicolumn{1}{|c|}{H$\cdot 10^{21}$} &\multirow{2}{*}{config}&\multicolumn{1}{c|}{$T_{obs}$}& \multirow{2}{*}{$N_c$}&\multirow{2}{*}{$\mathbf{\bar N}_b~^{5\%}_{95\%}$} & \multicolumn{2}{c|}{confidence interval} \\ 
\multicolumn{1}{|c|}{(Hz$^{-1}$)}& & \multicolumn{1}{c|}{(days)} & &  & 90\%& 95\%\\ \hline
\endhead
\hline
\multicolumn{7}{r}\textit{continue $\gg$ }
\endfoot
\hline
\hline
\endlastfoot
 \hline\hline 
\hline\hline 
 \multirow{1}{*}{2.00}
&	\textsc{au-na}&	4.2&	0&	0.4^{+0.034}_{-0.035}&	0\div 2.9&	0\div 3.6\\[1pt] \cline{2-7}\cline{2-7}
 \hline\hline 
 \multirow{1}{*}{2.24}
&	\textsc{au-na}&	6.9&	0&	0.4^{+0.033}_{-0.033}&	0\div 2.9&	0\div 3.6\\[1pt] \cline{2-7}\cline{2-7}
  \hline\hline 
 \multirow{1}{*}{2.51}
&	\textsc{au-na}&	10.9&	1&	0.5^{+0.036}_{-0.037}&	0\div 4.2&	0\div 5\\[1pt] \cline{2-7}\cline{2-7}
  \hline\hline 
 \multirow{1}{*}{2.82}
&	\textsc{al-na}&	53.7&	8&	7.9^{+0.15}_{-0.15}&	0\div 7.3&	0\div 8.6\\[1pt] \cline{2-7} 
 &	\textsc{al-au}&	38.9&	2&	2.9^{+0.089}_{-0.089}&	0\div 4.3&	0\div 5.2\\[1pt] \cline{2-7} 
 &	\textsc{au-na}&	12.0&	1&	0.6^{+0.04}_{-0.041}&	0\div 4.1&	0\div 5\\[1pt] \cline{2-7} 
 &	\textsc{al-au-na}&	5.9&	0&	4.9^{+3.4}_{-3.8} \cdot 10^{-4}&	0\div 2.9&	0\div 3.6\\[1pt] \cline{2-7} 
 &	\textsc{ex-na}&	1.7&	0&	0.2^{+0.023}_{-0.024}&	0\div 2.9&	0\div 3.6\\
\hline\hline 
 \end{longtable}
\end{squeezetable}

\begin{squeezetable}
\setlength\extrarowheight{2pt}
\setlength{\tabcolsep}{0.5pt}
\setlength{\LTcapwidth}{85mm}
\setcounter{LTchunksize}{50}
\begin{longtable}{|c|c|d|c|>{$}c<{$}|>{$}c<{$}|>{$}c<{$}|}
39.81 &	\textsc{al-au-ex}&	14.0&	0&	<9.1\cdot 10^{-5}&	0\div 2.9&	0\div 3.6 \kill
\caption{\label{tab:GC}Results of the event search analysis as described in the previous table but optimized for the Galactic Center Direction (see Section~\ref{sec:preprocessing}).}\\
\hline\hline 
\multicolumn{1}{|c|}{H$\cdot 10^{21}$} &\multirow{2}{*}{config}&\multicolumn{1}{c|}{$T_{obs}$}& \multirow{2}{*}{$N_c$}&\multirow{2}{*}{$\mathbf{\bar N}_b~^{5\%}_{95\%}$} & \multicolumn{2}{c|}{confidence interval} \\ 
\multicolumn{1}{|c|}{(Hz$^{-1}$)}& & \multicolumn{1}{c|}{(days)} & &  & 90\%& 95\%\\ \hline
\endfirsthead
\hline
\multicolumn{1}{|c|}{H$\cdot 10^{21}$} &\multirow{2}{*}{config}&\multicolumn{1}{c|}{$T_{obs}$}& \multirow{2}{*}{$N_c$}&\multirow{2}{*}{$\mathbf{\bar N}_b~^{5\%}_{95\%}$} & \multicolumn{2}{c|}{confidence interval} \\ 
\multicolumn{1}{|c|}{(Hz$^{-1}$)}& & \multicolumn{1}{c|}{(days)} & &  & 90\%& 95\%\\ \hline
\endhead
\hline
\multicolumn{7}{r}\textit{continue $\gg$ }
\endfoot
\hline
\hline
\endlastfoot
\hline\hline 
 \multirow{1}{*}{2.24}
&	\textsc{au-na}&	1.6&	0&	8.1^{+1.5}_{-1.5} \cdot 10^{-2}&	0\div 2.9&	0\div 3.6\\[1pt] \cline{2-7}\cline{2-7}
 \hline\hline 
 \multirow{1}{*}{2.51}
&	\textsc{au-na}&	3.1&	0&	0.1^{+0.017}_{-0.017}&	0\div 2.9&	0\div 3.6\\[1pt] \cline{2-7}\cline{2-7}
 \hline\hline 
 \multirow{2}{*}{2.82}
&	\textsc{al-na}&	5.9&	1&	0.9^{+0.05}_{-0.05}&	0\div 4&	0\div 4.8\\[1pt] \cline{2-7} 
 &	\textsc{al-au}&	5.0&	0&	0.4^{+0.032}_{-0.032}&	0\div 2.9&	0\div 3.6\\[1pt] \cline{2-7} 
 &	\textsc{au-na}&	3.2&	0&	0.2^{+0.025}_{-0.025}&	0\div 2.9&	0\div 3.6\\[1pt] \cline{2-7}\cline{2-7}
 &	\textit{total}&	14.0&	1&	1.5{\scriptstyle\pm 0.065}&	0\div 3.8&	0\div 4.6\\[1pt] \hline\hline 
 \multirow{2}{*}{3.16}
&	\textsc{al-na}&	61.3&	5&	3.6^{+0.098}_{-0.099}&	0\div 7&	0\div 8.1\\[1pt] \cline{2-7} 
 &	\textsc{al-au}&	36.3&	1&	1.5^{+0.063}_{-0.064}&	0\div 3.8&	0\div 4.6\\
\end{longtable}
\end{squeezetable}

\newpage 

\end{document}